\documentclass[iop]{emulateapj}
  
\usepackage{apjfonts}
\usepackage{amsmath}
\usepackage{epsf}
\usepackage{color}
\usepackage{comment}
\usepackage{amsmath}
\usepackage{graphicx}
\usepackage{subfigure}
\usepackage{epstopdf}
\usepackage{float}
\usepackage{enumitem}
\usepackage{appendix}

\shortauthors{Y.-K. Chiang et al.}

\begin{document}
\title{Surveying Galaxy Proto-clusters in Emission: A Large-scale Structure at \MakeLowercase{z}=2.44 and the Outlook for HETDEX}

\newcommand{\Mstar}{M_{\star}}
\newcommand{\Mbh}{M_{\rm BH}}
\newcommand{\Mhalo}{M_{\rm halo}}
\newcommand{\Muv}{M_{\rm UV}}
\newcommand{\Lam}{\Lambda}
\newcommand{\lam}{\lambda}
\newcommand{\Del}{\Delta}
\newcommand{\del}{\delta}
\newcommand{\mpc}{\rm Mpc}
\newcommand{\kpc}{\rm kpc}
\newcommand{\yr}{\rm yr}
\newcommand{\km}{\rm km}
\newcommand{\s}{\rm s}
\newcommand{\kms}{\rm km\,s^{-1}}
\newcommand{\erg}{\rm erg}
\newcommand{\Msun}{M_{\odot}}
\newcommand{\Lsun}{L_{\odot}}
\newcommand{\Zsun}{Z_{\odot}}
\newcommand{\hinv}{h^{-1}}
\newcommand{\himpc}{\hinv{\rm\,Mpc}}
\newcommand{\hikpc}{\hinv{\rm\,kpc}}
\newcommand{\himsun}{\,\hinv{\Msun}}

\newcommand{\Om}{\Omega_{\rm m}}
\newcommand{\Ol}{\Omega_{\Lam}}
\newcommand{\Ob}{\Omega_{\rm b}}
\newcommand{\OHI}{\Omega_{\rm HI}}
\newcommand{\HI}{H{\sc i}\,\,}
\newcommand{\NHI}{{N_{\rm HI}}}
\newcommand{\Mtot}{{M_{\rm tot}}}
\newcommand{\Lbox}{{L_{\rm box}}}
\newcommand{\Csto}{{C_{\rm stoc}}}
\newcommand{\Lya}{Ly$\alpha$}
\newcommand{\La}{L_{\rm Ly\alpha}}
\newcommand{\fLya}{f_{\rm Ly\alpha}}
\newcommand{\fesc}{f_{\rm esc}}
\newcommand{\fion}{f_{\rm esc}^{\rm ion}}
\newcommand{\figm}{f_{\rm IGM}}
\newcommand{\highz}{high-$z$}
\newcommand{\fdust}{f_{\rm dust}}
\newcommand{\SFR}{{\rm SFR}}
\newcommand{\lgZ}{\log (Z/Z_\odot)}

\author{Yi-Kuan Chiang\altaffilmark{1}, Roderik A. Overzier\altaffilmark{2}, Karl Gebhardt\altaffilmark{1}, Steven L. Finkelstein\altaffilmark{1}, Chi-Ting Chiang\altaffilmark{3}, Gary J. Hill\altaffilmark{4,1}, Guillermo A. Blanc\altaffilmark{5,6}, Niv Drory\altaffilmark{4}, Taylor S. Chonis\altaffilmark{1}, Gregory R. Zeimann\altaffilmark{7,8}, Alex Hagen\altaffilmark{7,8}, Donald P. Schneider\altaffilmark{7,8}, Shardha Jogee\altaffilmark{1}, Robin Ciardullo\altaffilmark{7,8}, and Caryl Gronwall\altaffilmark{7,8}}

\altaffiltext{1}{Department of Astronomy, University of Texas at Austin, 1 University Station C1400, Austin, TX 78712, USA}
\altaffiltext{2}{Observat\'orio Nacional, Rua Jos\'e Cristino, 77. CEP 20921-400, S\~ao Crist\'ov\~ao, Rio de Janeiro-RJ, Brazil}
\altaffiltext{3}{Max-Planck-Institut f\"ur Astrophysik, Karl-Schwarzschild-Str. 1, D-85741 Garching, Germany}
\altaffiltext{4}{McDonald Observatory, University of Texas at Austin, 1 University Station C1402, Austin, TX 78712, USA}
\altaffiltext{5}{Departamento de Astronom\'{\i}a, Universidad de Chile, Casilla 36-D, Santiago, Chile.}
\altaffiltext{6}{Observatories of the Carnegie Institution for Science, Pasadena, CA, USA}
\altaffiltext{7}{Department of Astronomy \& Astrophysics, The Pennsylvania State University, University Park, PA 16802, USA}
\altaffiltext{8}{Institute for Gravitation and the Cosmos, The Pennsylvania State University, University Park, PA 16802, USA}

\begin{abstract}

Galaxy proto-clusters at $z\gtrsim2$ provide a direct probe of the rapid mass assembly and galaxy growth of present day massive clusters. Because of the need of precise galaxy redshifts for density mapping and the prevalence of star formation before quenching, nearly all the proto-clusters known to date were confirmed by spectroscopy of galaxies with strong emission lines. Therefore, large emission-line galaxy surveys provide an efficient way to identify proto-clusters directly. Here we report the discovery of a large-scale structure at $z=2.44$ in the HETDEX Pilot Survey. On a scale of a few tens of Mpc comoving, this structure shows a complex overdensity of Ly$\alpha$ emitters (LAE), which coincides with broad-band selected galaxies in the COSMOS/UltraVISTA photometric and zCOSMOS spectroscopic catalogs, as well as overdensities of intergalactic gas revealed in the Ly$\alpha$ absorption maps of Lee et al. (2014). We construct mock LAE catalogs to predict the cosmic evolution of this structure. We find that such an overdensity should have already broken away from the Hubble flow, and part of the structure will collapse to form a galaxy cluster with $10^{14.5\pm0.4}$ $M_{\odot}$ by $z=0$. The structure contains a higher median stellar mass of broad-band selected galaxies, a boost of extended Ly$\alpha$ nebulae, and a marginal excess of active galactic nuclei relative to the field, supporting a scenario of accelerated galaxy evolution in cluster progenitors. Based on the correlation between galaxy overdensity and the $z=0$ descendant halo mass calibrated in the simulation, we predict that several hundred $1.9 < z < 3.5$ proto-clusters with $z = 0$ mass of $> 10^{14.5}$ $M_{\odot}$ will be discovered in the 8.5 Gpc$^3$ of space surveyed by the Hobby Eberly Telescope Dark Energy Experiment.

\end{abstract}
\keywords{cosmology: observations -- galaxies: clusters: general -- galaxies: evolution -- galaxies: high-redshift}

\section{Introduction}

Galaxy proto-clusters at $z\gtrsim2$ are the ``crime scene'' of the rapid mass assembly and galaxy growth of present day massive clusters. During this epoch, the most massive dark matter (DM) halos in cluster progenitors are just about to cross the characteristic mass scale of $10^{14}$ $M_{\odot}$ \citep{2013ApJ...779..127C,2013ApJ...763...70W}, coinciding with the increasing dominance of various intra-cluster processes seen in fully formed clusters. The total star formation rate (SFR) of a $z\gtrsim2$ proto-cluster is predicted to be $\sim3$ orders of magnitude higher than that of its $z=0$ descendant \citep{2013ApJ...770...57B}, implying a rapid build-up of the stellar content in line with an emerging quiescent galaxy population. Efficient baryon accretion of massive galaxies via cold streams from the gaseous cosmic web might be switching to an inefficient mode due to a uniformly shock-heated medium. Such a transition is expected to take place in the largest halos first, i.e., in cluster progenitors during this epoch \citep{2005MNRAS.363....2K,2006MNRAS.368....2D,2009Natur.457..451D}. The subsequent virialization on both galaxy and cluster scales in about a dynamical timescale largely erases the signatures of the aforementioned processes, placing a fundamental limit on inferences based on the largely archaeological record of cluster formation based upon near-field studies. Direct studies of cluster progenitors thus provide irreplaceable probes to understand the formation of present day massive clusters. 

The search for high-redshift cluster progenitors is challenging due to their lack of mature cluster signatures such as extended X-ray emission \citep{2011NJPh...13l5014F}, the Sunyaev--Zel'dovich effect \citep{2015ApJS..216...27B}, and the prominent galaxy red sequence \citep{2005ApJS..157....1G,2011AJ....141...94G}. The fundamental picture of gravitational structure formation implies that the most massive collapsed objects evolved from the densest regions in the early universe on a large scale \citep[][and references therein]{2012ARA&A..50..353K}. The finding of proto-clusters requires identifying galaxy overdensities in three-dimensions using precise redshift measurements \citep{2013ApJ...779..127C}.

Active star formation in cluster progenitors implies that (at least for the purpose of proto-cluster search and identification) more focus should be placed on star-forming galaxies instead of the quiescent ones that play a dominant role in traditional cluster studies. The difficulty in mapping the high-redshift cosmic density field is alleviated by the prevalence of emission lines in these star-forming galaxies, for which spectroscopic redshift can be obtained once the line transition is identified. Therefore, nearly all the $\sim25$ proto-clusters known to date \citep[see the recent compilation in][]{2013ApJ...779..127C} were found and/or confirmed spectroscopically by overdensities of galaxies with strong emission lines, particularly Ly$\alpha$ redshifted into the optical window \citep{1998ApJ...492..428S,2000ApJ...532..170S,2005ApJ...626...44S,2000A&A...358L...1K,2004A&A...428..793K,2000A&A...361L..25P,2002A&A...396..109P,2002ApJ...569L..11V,2004A&A...424L..17V,2005A&A...431..793V,2007A&A...461..823V,2003ApJ...586L.111S,2004ApJ...602..545P,2005ApJ...634L.125M,2005ApJ...620L...1O,2008ApJ...678L..77P,2011MNRAS.417.1088K,2012AJ....143...79Y,2014A&A...570A..16C,2014ApJ...796..126L,2014A&A...572A..41L,2015MNRAS.447.3069S}. Alternatively, H$\alpha$ emitters are also used as density tracers \citep{2011MNRAS.415.2993H,2011MNRAS.416.2041M,2012ApJ...757...15H,2013MNRAS.428.1551K}.

Massive proto-clusters at $z\gtrsim2$, although having a much larger radius of influence compared with clusters in the local universe, occupy only $\sim1/1000$ of the cosmic volume \citep{2013ApJ...779..127C}. Their abundance, by definition, is as low as that of galaxy clusters at $z=0$. An effective survey of proto-clusters thus needs to probe an extremely large volume. Traditional multi-object slit spectroscopy, although providing reliable redshifts and galaxy spectral diagnostics, is expensive as a survey tool of this scale. Narrow-band imaging (with a larger redshift uncertainty than direct spectroscopy) has been successful in finding overdensities of Ly$\alpha$ emitters (LAE) in both blank fields \citep{2005ApJ...620L...1O} and targeted fields around powerful radio galaxies \citep[see a summary in][]{2007A&A...461..823V}. This technique also revealed a puzzling but fascinating population of diffuse Ly$\alpha$ halos, the so called Ly$\alpha$ ``blobs'' in overdense regions \citep{2000ApJ...532..170S,2008ApJ...678L..77P,2009MNRAS.400L..66M,2009ApJ...693.1579Y,2011ApJ...740L..31E,2012MNRAS.425..878M}. However, narrow-band imaging typically requires a region of interest with a known redshift; if used as a survey tool, it probes only a small volume in a thin redshift slice of $\Delta z \sim 0.1$.

Blind spectroscopy provides an opportunity to largely increase the survey volume. For instance, wide-field slitless grism or prism spectroscopy (e.g., the baseline redshift surveys of the future \textit{Wide Field Infrared Survey Telescope} \citep[WFIRST;][]{2013arXiv1305.5422S,2015arXiv150303757S} and the \textit{Euclid} mission \citep{2011arXiv1110.3193L}) is particularly suitable for the searches of proto-clusters traced by bright emission-line galaxies. Integral field unit (IFU) spectroscopy has even greater potential, with no trade-off between spectral resolution and the survey depth due to spectral crowding and confusion (compared to grism surveys). For the same reason of source crowding, blind grism spectroscopy strongly demands space-based spatial resolution, while the IFU technique is feasible with ground-based facilities. However, early IFU techniques have focused on achieving sub-arcsecond sampling in a relatively small field of view \citep[e.g.,][]{2003SPIE.4841.1548E,2006SPIE.6269E..1AL,2010SPIE.7735E..08B,2015A&A...575A..75B}, making them less suitable for proto-cluster searches.

The Hobby Eberly Telescope Dark Energy Experiment \citep[HETDEX;][]{2008ASPC..399..115H} is pioneering the instrumentation development and observations of high-redshift large-scale structures using wide-field IFUs. In a 3 year baseline starting from late 2015, HETDEX will leverage the cosmic evolution of the dark energy equation of state with high-redshift ($z>2$) constraints imprinted by the Baryon Acoustic Oscillations \citep[BAO;][]{2005NewAR..49..360E} in the early universe. The program will perform a redshift survey of LAEs in 300 deg$^2$ (Spring field) plus 150 deg$^2$ (Fall field) at $1.9<z<3.5$ (with a filling factor of 1/4.5), with a total survey volume of $\sim8.5$ Gpc$^3$. The survey uses the 10-m Hobby--Eberly Telescope \citep[HET;][]{1998SPIE.3352...34R} with a wide-field upgrade to reach a $22\times22$ arcmin$^2$ field of view. Blind spectroscopy ($R\sim750$ in 350--550\,nm) with no pre-selection of targets will be performed using the Visible Integral-field Replicable Unit Spectrograph \citep[VIRUS;][]{2012SPIE.8446E..0NH,2014SPIE.9147E..0QH}. With the LAE redshifts, HETDEX will pinpoint numerous locations of the highest density concentrations at $1.9<z<3.5$, generating a substantially large and homogeneous sample of cluster progenitors in the key epoch of cluster formation before virialization. 

As a proof of concept, the HETDEX Pilot Survey \citep[HPS;][]{2011ApJS..192....5A} performed blind spectroscopy over a 169 arcmin$^2$ area (divided into four sub-fields) for bright LAEs at $1.9 < z < 3.8$, which corresponds to a volume of $\sim 10^6$ Mpc$^3$. A total of 105 LAEs were discovered and studied in detail \citep{2011ApJS..192....5A,2011ApJ...736...31B,2011ApJ...729..140F,2013ApJ...775...99C,2014ApJ...786...59H,2014ApJ...791....3S}.

Among the LAEs discovered in HPS, there is a concentration of nine LAEs across a 71.6 arcmin$^2$ region in the HPS-COSMOS field, which lie in a narrow redshift range at $z\sim2.44$ (LAE overdensity of $\gtrsim4$ in a comoving volume of $\sim 10 \times 10 \times 35$ $\rm{Mpc^3\ h^{-3}}$). Here we present a detailed characterization of this structure using HPS data, supplemented with a publicly available catalog of continuum-selected galaxies with photometric redshifts from COSMOS/UltraVISTA. We use a cosmological simulation to model the realistic connection between LAEs and the underlying matter field and the complex nonlinear gravitational structure formation across cosmic history. Our study shows that part of this structure will collapse to form a galaxy cluster with $10^{14.5 \pm 0.4}$ $M_{\odot}$ by $z = 0$. The structure (together with another similar overdensity partially covered by HPS) hosts several extended Ly$\alpha$ halos, some of which are identified as active galactic nucleus (AGN) in the X-ray. These systems are commonly found in overdense regions at high redshift, perhaps indicating an accelerated co-evolution of massive galaxies and their supermassive black holes in overdense environments.

In \S\,2, we describe our LAEs and continuum-selected galaxy sample, and we construct a suite of mock LAE catalogs with clustering properties bracketing that of the observed LAEs. In \S\,3, we present the spatial distributions of galaxies in HPS-COSMOS along the line of sight and on the projected sky. In \S\,4, we place this structure in the context of cosmic structure formation based on the cosmological simulation connected through the mocks. In \S\,5, we demonstrate a significant enhancement of diffuse Ly$\alpha$ halos and AGN in this structure. In \S\,6, we present the outlook for proto-cluster identification in the HETDEX survey. We discuss the results in \S\,7 and conclude this work in \S\,8. Cosmological parameters based on the 7 year Wilkinson Microwave Anisotropy Probe \citep{2011ApJS..192...18K} are adopted: [\,$h$, $\Omega_m$, $\Omega_{\Lambda}$, $n_s$, $\sigma_8\,]\,=\,[\,$0.704, 0.272, 0.728, 0.967, 0.81]. All magnitudes given are in the AB system \citep{1983ApJ...266..713O}.

\section{Galaxy Samples and Simulations}
Cluster formation is directly driven by the evolution of the matter density field under gravitational processes. However, DM, being the dominant component of the matter density, has no direct electromagnetic signature. We follow the standard formalism using galaxies as (in general, biased) tracers of the underlying density field. Here we describe our HPS LAE sample and the COSMOS/UltraVISTA catalog of continuum-selected (``photo-$z$'') galaxies. We generate mock LAEs matched in bias and stellar mass to bridge the gaps between luminous and DM, and also connect observations in a fixed lightcone to cosmological simulations that model their time evolution.

\subsection{L\MakeLowercase{y}$\alpha$ Emitters: The HETDEX Pilot Survey}

In this work, we use mainly the LAE sample in HPS-COSMOS, the largest contiguous HPS sub-field of 71.6 arcmin$^2$ ($\sim 7' \times 10'$) near the center of the COSMOS field. The sample contains a total of 52 LAEs at $1.9 < z < 3.8$, with four showing X-ray emission \citep[matched with the catalog of][]{2009ApJS..184..158E}\footnote{In this work we do not exclude AGNs from the LAE sample since all the Ly$\alpha$ emitting objects provide reliable redshifts to trace the underlying cosmic density field. We model the clustering properties of the full LAE population in \S\,2.3. This treatment is favored for future applications of the full HETDEX survey, in which no coordinated deep X-ray observations are planned to cover a significant fraction of the wide HETDEX field.}. The field of HPS-COSMOS partially overlaps with several deep surveys that cover the redshift of $>2$ including CANDELS (PI: Faber, Ferguson), VVDS/VUDS (PI: Le F{\`e}vre), zCOSMOS (PI: Lilly), ZFOURGE (PI: Labb{\'e}), 3D-HST (PI: van Dokkum), and a pilot survey of CLAMATO (PI: Lee) for Ly$\alpha$ forest tomography. We will refer to some of the findings from these surveys when relevant.

HPS \citep{2011ApJS..192....5A} is a blind spectroscopic survey of emission-line galaxies using the Mitchell Spectrograph, formerly called the VIRUS-P spectrograph \citep[the VIRUS prototype;][]{2008SPIE.7014E..70H} on the 2.7 m Harlan J. Smith telescope at McDonald Observatory. A single Mitchell Spectrograph pointing covers an area of $1.7'\times1.7'$ with a 1/3 filling factor using an array of 246 fibers of each $4''.235$ in diameter. With a 6-dither pattern, HPS reaches a complete coverage in the field and sub-fiber-size spatial sampling. The survey contains four sub-fields in COSMOS (71.6 arcmin$^2$), GOODS-N (35.5 arcmin$^2$), MUNICS (49.9 arcmin$^2$), and XMM-LSS (12.3 arcmin$^2$) that are rich in ancillary multi-wavelength data, with a total survey area of 169 arcmin$^2$. The spectra cover a bandpass of 3500--5800 \AA\ with a spectral FWHM of 5 \AA\ ($\sigma_{inst}\sim130$ km s$^{-1}$ at 5000 \AA). The survey probes LAEs at $1.9 < z < 3.8$ with a single line expected within the bandpass in a total effective volume of $\sim 10^6$ Mpc$^3$. Each line detection is matched with a continuum counterpart or an upper limit is determined if undetected. LAEs are then distinguished from lower redshift galaxies with a single line detection (mainly unresolved [O II]$\lambda$3727, 3729 emitters at $0.19<z<0.56$) by an equivalent width (EW) criterion, where objects with a rest-frame EW$_{\rm Ly\alpha} > 20$ \AA\ are classified as LAEs. The contamination rate is estimated to be 4\%--10\%. A total of 105 LAEs are discovered in HPS down to a L$_{\rm Ly\alpha}^{\rm obs}$ limit of $\sim 4 \times 10^{42}$ erg s$^{-1}$ (roughly constant across the redshift range). Six of the 105 LAEs have X-ray counterparts, indicating the presence of AGNs in $\sim5\%$ of the sample. Among the nine LAEs in the large-scale structure at $z=2.44$ (see \S\,3 and Table 1), four are covered by 3D-HST. The HPS LAE identifications for these four sources are all confirmed by at least one 3D-HST metal line detection. An additional two LAEs (and also the four covered by 3D-HST) in the $z=2.44$ structure are followed up and confirmed using Magellan/IMACS spectroscopy with a spectral resolution of 150 km s$^{-1}$ FWHM, revealing unique asymmetric line profiles expected for Ly$\alpha$, and excluding the possibilities of being foreground [O II] emitters of $\lambda$3727, 3729 doublet (Chonis et al. in prep.).

\cite{2014ApJ...786...59H} estimated the stellar mass of 63 out of the total 74 LAEs in the HPS GOODS-N and COSMOS fields by spectral energy distribution (SED) fitting of individual galaxies, finding a wide distribution of ${\rm log} \, (M_* / M_{\odot})$ spanning from $\sim 7.5$ to $\sim 10.5$.

\def\arraystretch{1.5}
\begin{centering}
\begin{deluxetable*}{cccccccccccc}
\tablecaption{HPS-COSMOS $\rm{Ly\alpha}$ Emitter Catalog (Selected)} \tablewidth{0pt}
\tablehead{	\colhead{HPS}	&
		\colhead{$z$\tablenotemark{a}}	&
		\colhead{$\alpha$}	&
		\colhead{$\delta$}	&
		\colhead{Flux}	&
		\colhead{L}	&
		\colhead{Spectral}	&
		\colhead{Spatial}	&
		\colhead{Counter-} &
		\colhead{Counter-} &
		\colhead{EW$_{\rm{rest}}$$\rm$\tablenotemark{e}}	&
		\colhead{Flux$\rm{_{X-ray}}$}	\\
		\colhead{Index}	&
		\colhead{($\rm{Ly\alpha}$)}	&
		\colhead{(J2000)}	&
		\colhead{(J2000)}	&
		\colhead{($\rm{Ly\alpha}$)}	&
		\colhead{($\rm{Ly\alpha}$)}	&
		\colhead{FWHM$\rm$\tablenotemark{b}}	&
		\colhead{FWHM$\rm$\tablenotemark{c}}	&
		\colhead{part m$\rm{_R}$}	&
		\colhead{part Prob.$\rm$\tablenotemark{d}}	&
		\colhead{($\rm{Ly\alpha}$)}	&
		\colhead{(0.5--10 keV)}	\\
		\colhead{}	&
		\colhead{}	&
		\colhead{[deg]}	&
		\colhead{[deg]}	&
		\colhead{[$10^{-17}$ cgs]}	&
		\colhead{[$10^{42}$ cgs]}	&
		\colhead{[km s$^{-1}$]}	&
		\colhead{[arcsec]}	&
		\colhead{[mag]}	&
		\colhead{}	&
		\colhead{[\AA]}	&
		\colhead{[$10^{-17}$ cgs]}}
\startdata
\noalign{\vskip -1mm} 
\multicolumn{12}{c}{\bf{$\bf{z=2.44}$ structure}}\\
\noalign{\vskip 1.5mm} 
160 & 2.4346 & 150.03587 & 2.29406 & 17.1$^{+10.5}_{-6.4}$ & 8.3$^{+5.1}_{-3.1}$ & 663 & 5.2$^{+1.5}_{-1.6}$ & 27.35 & 0.61 & 1034.3$^{+1000.0}_{-559.0}$ &  \\
162 & 2.4284 & 150.03637 & 2.25889 & 76.4$^{+14.6}_{-11.5}$ & 37.0$^{+7.1}_{-5.6}$ & 1063 & 8.3$^{+1.3}_{-0.9}$ & 24.45 & 0.20 & 564.3$^{+165.4}_{-114.8}$ & 370$\pm67$ \\
164 & 2.4518 & 150.03729 & 2.28978 & 25.4$^{+13.7}_{-12.9}$ & 12.6$^{+6.8}_{-6.4}$ & 482 & 11.0$^{+3.3}_{-3.3}$ & 24.32 & 0.31 & 126.4$^{+70.5}_{-64.5}$ &  \\
182 & 2.4337 & 150.05137 & 2.23778 & 25.6$^{+5.8}_{-5.2}$ & 12.5$^{+2.8}_{-2.5}$ & 211 & 4.9$^{+0.5}_{-0.8}$ & 25.04 & 0.60 & 180.8$^{+49.4}_{-40.5}$ &  \\
189 & 2.4515 & 150.05462 & 2.31564 & 12.9$^{+8.7}_{-6.7}$ & 6.4$^{+4.3}_{-3.3}$ & 509 & 5.1$^{+1.8}_{-1.9}$ & 24.99 & 0.64 & 85.3$^{+59.2}_{-44.6}$ &  \\
197 & 2.4419 & 150.06121 & 2.29650 & 17.8$^{+7.1}_{-6.0}$ & 8.7$^{+3.5}_{-2.9}$ & 536 & 4.1$^{+1.2}_{-1.2}$ & 25.8 & 0.33 & 258.9$^{+317.6}_{-114.9}$ &  \\
263 & 2.4323 & 150.12108 & 2.23589 & 24.1$^{+8.0}_{-7.7}$ & 11.7$^{+3.9}_{-3.7}$ & 511 & 5.8$^{+1.2}_{-1.0}$ & 24.17 & 0.89 & 66.3$^{+22.9}_{-21.4}$ &  \\
306 & 2.4390 & 150.16504 & 2.22739 & 38.3$^{+5.8}_{-9.2}$ & 18.7$^{+2.8}_{-4.5}$ & 766 & 7.1$^{+1.0}_{-0.8}$ & 24.07 & 0.72 & 90.4$^{+15.6}_{-22.2}$ &  \\
318 & 2.4558 & 150.18387 & 2.26636 & 30.3$^{+8.9}_{-11.1}$ & 15.1$^{+4.4}_{-5.5}$ & 349 & 8.0$^{+1.7}_{-1.6}$ & 23.69 & 0.32 & 74.5$^{+22.9}_{-27.6}$ &  \\
\hline
\noalign{\vskip 1.5mm} 
\multicolumn{12}{c}{\bf{Other Extended LAEs or AGNs}}\\
\noalign{\vskip 1.5mm} 
145 & 2.1751 & 150.02608 & 2.21969 & 84.0$^{+14.8}_{-8.1}$ & 31.0$^{+5.5}_{-3.0}$ & 1164 & 7.5$^{+0.8}_{-0.8}$ & 24.08 & 0.51 & 2380.9$^{+1000.0}_{-1190.5}$ &  \\
148 & 3.4176 & 150.02917 & 2.32439 & 8.6$^{+2.0}_{-2.6}$ & 9.5$^{+2.2}_{-2.9}$ & 289 & 4.5$^{+1.2}_{-1.0}$ & 24.77 & 0.43 & 180.5$^{+74.1}_{-62.4}$ & 166$\pm49$ \\
222 & 2.9430 & 150.07600 & 2.26417 & 87.3$^{+4.7}_{-5.8}$ & 67.5$^{+3.6}_{-4.5}$ & 983 & 4.7$^{+0.2}_{-0.2}$ & 23.55 & 0.98 & 278.1$^{+41.0}_{-34.5}$ & 268$\pm60$ \\
261 & 2.0960 & 150.11904 & 2.29678 & 143.7$^{+23.2}_{-10.1}$ & 48.4$^{+7.8}_{-3.4}$ & 886 & 8.3$^{+0.9}_{-0.6}$ & 23.76 & 0.87 & 536.7$^{+157.8}_{-92.4}$ & 2040$\pm125$ \\
\noalign{\vskip -3mm}
\enddata
\tablenotetext{a}{With an uncertainty of $4\times 10^{-4}$ based on a 0.5 \AA\ line center uncertainty.}
\tablenotetext{b}{After deconvolution with a 5 \AA\ FWHM instrumental resolution ($\sigma_{inst}\sim130$ km s$^{-1}$).}
\tablenotetext{c}{Including a tophat component of the fiber size of $4''.235$ and the effects of dither pattern and discrete sampling.}
\tablenotetext{d}{Probability of counterpart association ($R$-band).}
\tablenotetext{e}{Based on an interpolation between the two nearest filters for continuum.}

\end{deluxetable*}
\end{centering}

\cite{2011ApJS..192....5A} performed a curve of growth analysis to obtain robust total Ly$\alpha$ fluxes for the HPS LAEs and estimated the spatial extent of Ly$\alpha$ emission. As the survey uses $4''.235$ diameter fibers with a dither pattern to achieve a discrete sampling of $\lesssim3''$ nearest fiber-center distances, no constraint below the scale of few arcsecond is obtained. Nonetheless, sources with an apparent spatial $\rm FWHM > 6''.81$ (including the effects of instrument, sampling, and seeing) can be ruled out as point sources with a confidence level of $99.7\%$\footnote{The scale of $\sim7''$ coincides with the sum of the fiber size and the average sampling separation. A source of $\sim7''$ would be detected (for each at least about half fiber area is filled) by 10--12 fibers, while a source of $6''$ would be detected by only 4--6 fibers \citep[see Figure 1 in][]{2011ApJS..192....5A}.}. Using this criterion, there are a total of 7 (10) extended Ly$\alpha$ halos in HPS-COSMOS (full HPS). Table 1 presents the catalog for a selected subset of LAEs of interest in the HPS-COSMOS.

\subsection{Continuum-selected Photo-z Galaxies: The COSMOS/UltraVISTA}

We supplement the LAEs with continuum-selected galaxies with photometric redshifts (photo-$z$). Although their redshift uncertainty is considerably larger than that of the LAEs, these objects provide a more mass complete sample and over a wider field. We use a publicly available $K_s$ band selected photometric redshift galaxy catalog of \cite{2013ApJS..206....8M} in the 1.62 deg$^2$ COSMOS/UltraVISTA survey. The catalog combines photometric datasets from UltraVISTA \citep{2012A&A...544A.156M} for near-IR, Subaru/SuprimeCam \citep{2007ApJS..172....9T} and CFHT/MegaCam \citep{2007ApJS..172...99C} for optical. Information from the GALEX FUV and NUV \citep{2005ApJ...619L...1M} and $Spitzer$ IRAC+MIPS mid-IR data \citep{2007ApJS..172...86S} are included. The photo-$z$ error of galaxies at $2<z<3$ is, on average, at a level of $\sigma_z/(1+z) = 2.5$--$3\%$. Here we use the sample above the 90\% completeness limit of $K_s < 23.4$ mag, excluding a small fraction ($\sim 4\%$) of galaxies showing a broad and/or multi-modal redshift probability distribution. A subsample of $K_s < 22.0$ galaxies will be referred to as the ``bright'' sample.

We will also use the galaxy stellar masses provided in \cite{2013ApJS..206....8M}, derived by SED fitting with the FAST code \citep{2009ApJ...700..221K} using a set of population synthesis models from \cite{2003MNRAS.344.1000B}. Solar metallicity, a \cite{2003PASP..115..763C} initial mass function (IMF), and a \cite{2000ApJ...533..682C} dust extinction law are assumed. The uncertainty in stellar mass is $\sim 0.2$ dex.

\subsection{Bias and Mass Matched Catalogs of Mock LAE}

$\Lambda$CDM cosmological $N$-body simulations and semi-analytic models (SAM) of galaxy formation provide a framework to model the complex hierarchical growth of DM and galaxies in three-dimensions on the relevant scales, and link the evolution of large-scale structures across cosmic time. To characterize the LAE density concentration in HPS-COSMOS at $z=2.44$ (see \S\,3), we generate a set of mock catalogs of LAEs at $z\sim2.4$ with realistic clustering properties by post-processing the SAM of \cite{2013MNRAS.428.1351G} on top of a new run of the Millennium Run (MR) cosmological DM N-body simulation \citep{2005Natur.435..629S} with the WMAP7 cosmology \citep{2011ApJS..192...18K}. The \cite{2013MNRAS.428.1351G} model improves upon the extensively tested models of \cite{2007MNRAS.375....2D} and \cite{2011MNRAS.413..101G}. Various galaxy properties are reasonably reproduced, and we particularly rely on its agreement with observations for galaxy clustering on large-scales in the ``two-halo'' regime \citep{2009MNRAS.396...39G,2011MNRAS.413..101G,2013MNRAS.428.1351G,2012MNRAS.422..804K,2013A&A...557A..17M,2014ApJ...782L...3C,2014MNRAS.437.3385K,2014MNRAS.442.1930P,2014ApJ...784..128S}. The galaxy stellar mass is 95\% and 60\% complete to $10^8$ $M_{\odot}$ and $10^7$ $M_{\odot}$,\footnote{The mass completeness is evaluated by comparing with the same galaxy model applied to the Millennium-II Simulation \citep[][]{2009MNRAS.398.1150B} with a higher mass resolution.} sufficient for the LAE modeling here. 

We aim to match simultaneously the LAE number density, the galaxy bias, and the stellar mass distribution to the observed sample. Correlation length analyses suggest that high-redshift LAEs are less clustered compared to broad-band selected Lyman-break galaxies (of a typical limiting magnitude of $K < 23$), with an overall linear bias of $2.0\pm 0.6$ at $2\lesssim z \lesssim 3$ \citep{2007ApJ...671..278G,2008ApJS..176..301O,2010ApJ...714..255G,2010ApJ...723..869O,2015arXiv150101215B}, and 2.5--4 at $z\sim4$ \citep{2007ApJ...668...15K,2008ApJS..176..301O,2010ApJ...723..869O}. Galaxy bias is known to correlate strongly with stellar mass \citep[or color/bolometric luminosity, see][and references therein]{2013pss6.book..387C} but very weakly with Ly$\alpha$ luminosity \citep{2008MNRAS.391.1589O}. Therefore the criterion to match the distribution in stellar mass provides constraints on not only the effective galaxy bias of the entire population but also the scatter of the bias (i.e., cosmic variance of the galaxy bias). The effect of the latter cannot be neglected in the case of localized statistics in real space, which usually suffer from having a relatively small number of objects. Conversely, the effect is less important in global statistics of correlation function and power spectrum.
A full theoretical modeling of Ly$\alpha$ radiative processes and a detailed match in Ly$\alpha$ luminosity function and EW distribution are not required since these have negligible impact on the gravitational clustering of LAEs once the criteria in bias and stellar mass are met.

To test the effects of clustering modeling on the final interpretation of the observed structure, we generate a suite of mock catalogs with four different galaxy bias and stellar mass distributions varied continuously to bracket that estimated for the observed LAEs. For each mock catalog, we artificially elevate the SFR of SAM galaxies, such that the same observational selection criteria of Ly$\alpha$ emission propagate to selecting model galaxies with different clustering via the positive correlation of SFR versus $M_*$, the ``star-forming main sequence'' \citep[e.g.,][]{2012ApJ...754...25R,2014MNRAS.443...19R}. This treatment acknowledges the uncertainties and bypasses the issue that at $z\sim2$, most of the state-of-the-art cosmological simulations (both hydrodynamical and SAM, including the one used here) generate a star-forming main sequence with a normalization 0.1--0.4 dex lower than that observed \citep[][and references therein]{2014ApJS..214...15S}, and not sufficiently ``bursty'' across the full range of stellar mass\footnote{The deficit of star-bursting objects in simulations across the star-forming sequence at $z\sim2$ results in a situation that only massive objects (thus high SFR) with low dust content would reach a high Ly$\alpha$ luminosity and EW. Such a population is significantly more massive than that observed. Thus we assume that the ranks in SFR for objects with a given stellar mass are statistically realistic in the simulation, but a large fraction of objects should have a higher absolute value of SFR. We will show later that by implementing a systematic SFR offset, other major galaxy properties of interest can be self-consistently reproduced and matched with that observed. This result indicates that the discrepancy in the normalization of the star-forming sequence is the sole fundamental problem at the level relevant to this work, which needs to be resolved in future SAMs.} \citep[see discussions in][]{2012MNRAS.426.2797W,2014arXiv1410.3485F,2014MNRAS.445..175G,2014MNRAS.444.2637M,2015ApJ...799..201W}.

A brief outline of our LAE modeling is as follows. We first systematically ``burst'' the SFR of SAM galaxies on the star-forming sequence. We then model the intrinsic Ly$\alpha$ production by galaxy instantaneous SFR, and the effects of dust attenuation by empirical constraints, effectively generating a broad distribution in galaxy $M_*$ that is consistent with what is found in observations. A high degree of stochasticity \citep[a survival probability, equivalently a Ly$\alpha$ duty cycle; see also][]{2010PASJ...62.1455N} is then adjusted by hand to match the observed HPS LAE number density in each mock catalog. Finally, an evaluation of the two-point correlation function of the mocks is preformed, which serves as a check of the Ly$\alpha$ modeling and of this approach as a tool to study large-scale structure. A summary of the suite of four mock LAE catalogs is given in Table 2. We describe the details of these procedures in the following.

After applying the SFR offset, we first compute, for each SAM galaxy, the intrinsic Ly$\alpha$ luminosity $L\rm{_{Ly\alpha}^{int}}$ generated in star-forming HII regions using the empirical calibration for H$\alpha$ \citep{1998ARA&A..36..189K} and assuming an intrinsic Ly$\alpha$ to H$\alpha$ ratio under Case B recombination \citep{1971MNRAS.153..471B,2006agna.book.....O}. This gives
\begin{equation}
\La^{\rm int} = 1.98 \times 10^{42} \, (\SFR / \Msun \, \yr^{-1}) \, \erg \, s^{-1},
\end{equation}
where the proportionality constant has been multiplied by a factor of 1.8 to convert from the Salpeter IMF \citep[][]{1955ApJ...121..161S} to Chabrier IMF \citep{2003PASP..115..763C} assumed in the SAM used here (both in a range of 0.1--100 M$_{\odot}$).

\def\arraystretch{1.5}
\begin{centering}
\begin{deluxetable}{lcccc}
\tablecaption{Summary of the Mock LAE Catalogs at $z=2.4$} \tablewidth{0pt}
\tablehead{	\colhead{Simulation}	&
		\colhead{$\Delta {\rm log \, (SFR/ \Msun \, \yr^{-1})}$\tablenotemark{a}}	&
		\colhead{$b$\tablenotemark{b}}	&
		\colhead{${\rm log} \, (M_* / M_{\odot})$$\rm$\tablenotemark{c}} &
		\colhead{$\rm P_{\rm LAE}$\tablenotemark{d}}}
\startdata
Mock I &  1.0 & 1.82 & $8.68^{+0.72}_{-0.65}$ &  3\% \\
\bf{Mock II} &  \bf{0.7} & \bf{2.00} & $\bf{8.99^{+0.61}_{-0.52}}$ &  \bf{4\%} \\
Mock III &  0.4 & 2.22 & $9.28^{+0.52}_{-0.44}$ &  6\% \\
Mock IV &  0.2 & 2.44 & $9.46^{+0.46}_{-0.39}$ & 9\%
\enddata
\tablenotetext{a}{The $\rm log \, (SFR/ \Msun \, \yr^{-1})$ offset applied to all galaxies in a given mock to compensate the systematically low and insufficiently bursty SFR of the SAM at this redshift. This offset is used as the sole control variable, which generates mocks with different bias and stellar mass.}
\tablenotetext{b}{Galaxy bias calculated at 8 Mpc h$^{-1}$ comoving.}
\tablenotetext{c}{Median and 16/84 percentiles of the stellar mass distribution. For comparison, the HPS star-forming LAEs at $1.9<z<3.8$ are estimated to have ${\rm log} \, (M_* / M_{\odot}) = 8.74^{+0.61}_{-0.71}$ \citep[converted to Chabrier IMF and the cosmological parameters adopted here;][]{2014ApJ...786...59H}.}
\tablenotetext{d}{Probability for star-forming galaxies to have the maximum values of $f_{\rm Ly\alpha}^{\rm \, esc}$ (Equation (2)), or equivalently the Ly$\alpha$ duty cycle, tuned to match the observed LAE number density.}
\end{deluxetable}
\end{centering}

Next, we implement dust attenuation of Ly$\alpha$ photons in the host galaxies. Due to the resonant nature of the transition, Ly$\alpha$ photons could experience long scattering path-lengths in the neutral interstellar medium (ISM) of the host galaxies. Thus a small amount of dust often produces a significant level of absorption. As a result, only several percent of the full star-forming galaxy population emit observable Ly$\alpha$ emission \citep{2010Natur.464..562H,2014ApJ...796...64C}. However, objects observed as LAEs show a level of dust attenuation in Ly$\alpha$ roughly following that of the stellar continuum at 1216 \AA\ \citep{2009ApJ...691..465F,2011ApJ...729..140F,2011ApJ...736...31B,2012ApJ...745...12N,2014ApJ...786...59H}. Specifically, observed LAEs show a lower limit of Ly$\alpha$ optical depth $\tau_{\rm Ly\alpha}$ that roughly equals to $\tau_{1216}$, the optical depth of stellar continuum at 1216 \AA, causing an upper limit in Ly$\alpha$ escape fraction,
\begin{equation}
\begin{split}
max\,(\,f_{\rm Ly\alpha}^{\rm \, esc}\,) = 10^{-0.4 \, k_{1216} \, E(B-V)} = 10^{-4.79 \, E(B-V)},
\end{split}
\end{equation}
assuming the extinction of the stellar continuum follows the \cite{2000ApJ...533..682C} law ($k_{1216}=11.98$). On the other hand, the observational criteria (mainly EW, but also $\La^{\rm obs}$) introduce a selection effect such that galaxies with $\tau_{\rm Ly\alpha} >> \tau_{1216}$ would not be observed as LAEs. Thus we set a Ly$\alpha$ escape fraction $f_{\rm Ly\alpha}^{\rm \, esc}$ at the maximum values with the given dust content of a galaxy, selecting those passing the $\La^{\rm obs}$ threshold of the HPS, and then drop a large fraction of galaxies according to a survival probability (independent of any galaxy properties) to match the observed LAE number density.\footnote{A detailed matching of the simulated and observed Ly$\alpha$ luminosity function and EW distribution requires relaxing our simplistic assumption for $f_{\rm Ly\alpha}^{\rm \, esc}$. As long as the clustering properties of LAEs are reproduced, however, we do not perform a fine-tuning in the probability distribution of $f_{\rm Ly\alpha}^{\rm \, esc}$, which in principle might correlate with environment and multiple bulk and unresolved galaxy properties other than the total amount of dust.} The dropped population corresponds to the large ($>90\%$) fraction of star-forming galaxies with $\tau_{\rm Ly\alpha} >> \tau_{1216}$, thus produces no observable Ly$\alpha$ emission.

\begin{figure}[t]
	\label{mock_LAEs}
     \begin{center}
        \hspace*{-0.2em}
        \subfigure{%
           \includegraphics[width=0.47\textwidth]{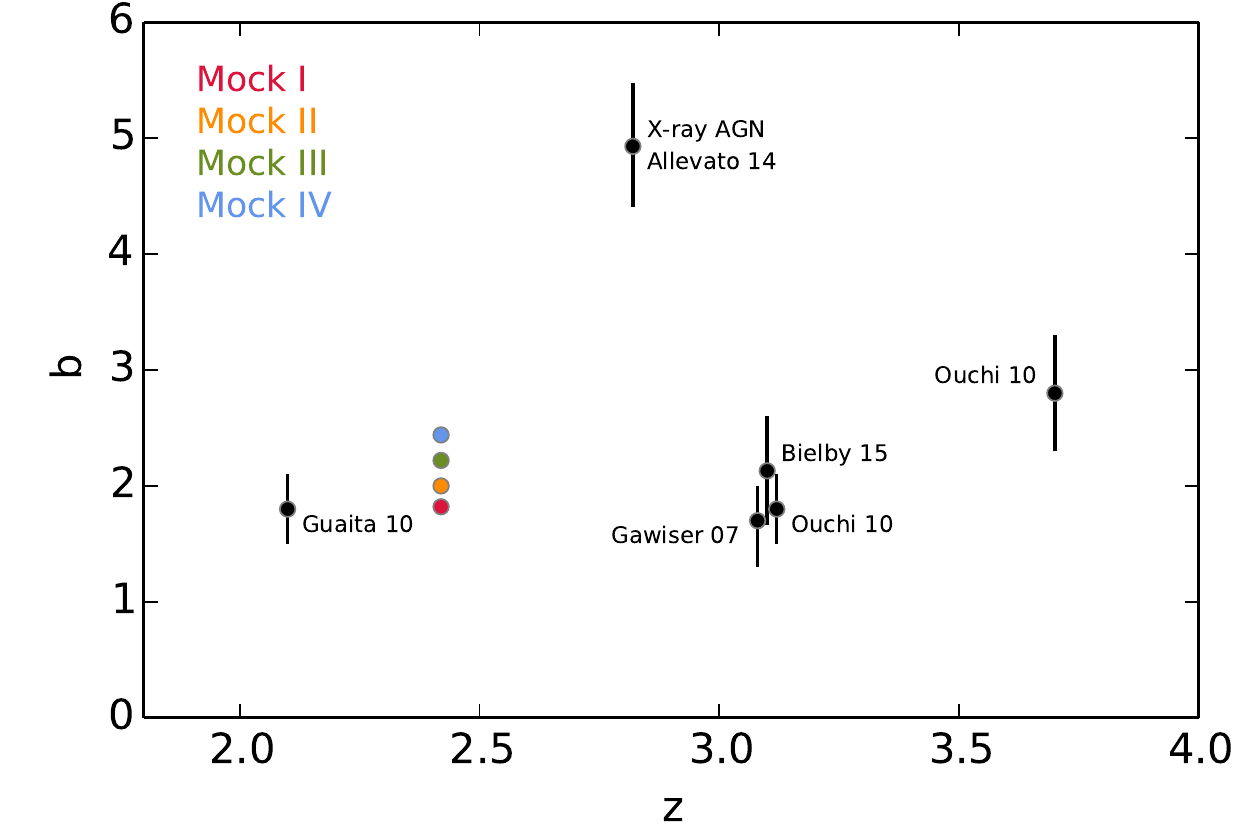}
        }\\ 
        \vspace*{0.5em}
        \hspace*{-0.2em}
        \subfigure{%
           \includegraphics[width=0.47\textwidth]{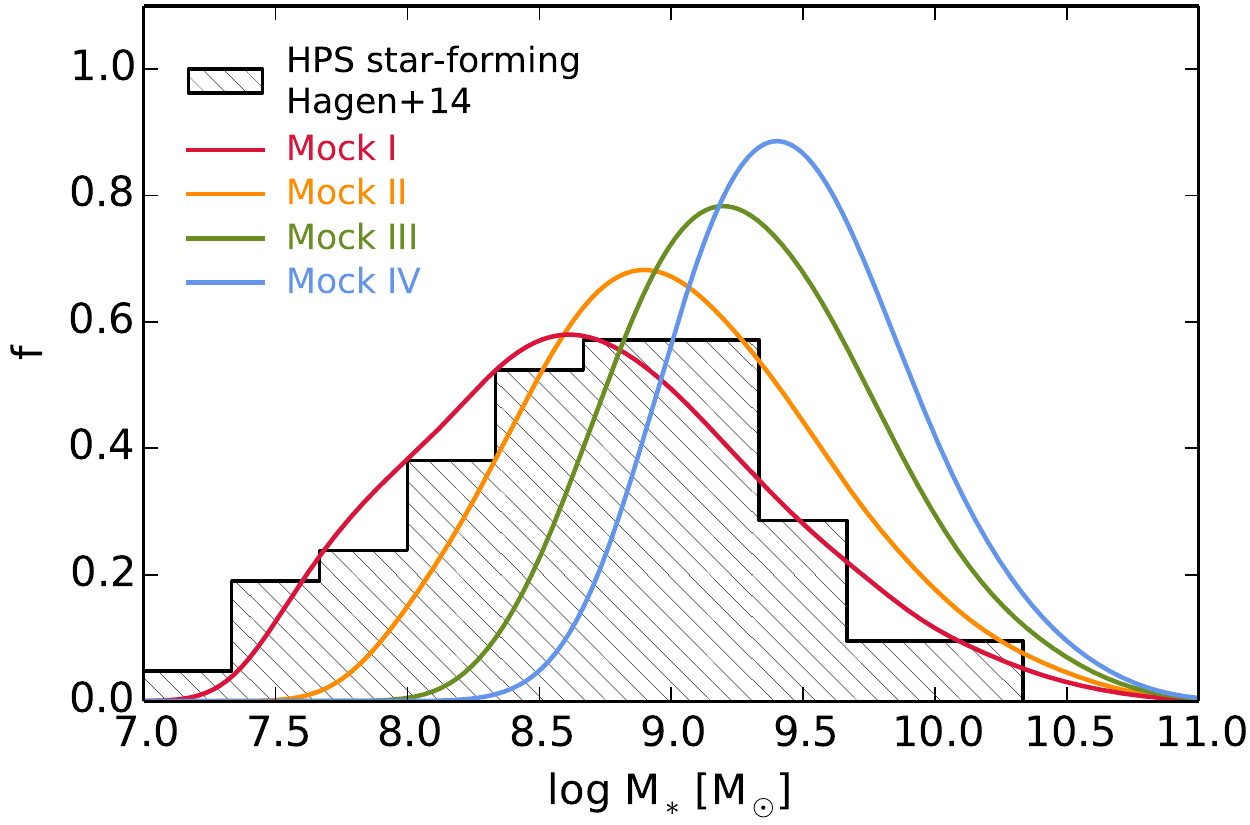}
        }\\ 
        \vspace*{-0.8em}       
    \end{center}
    \caption{%
    The galaxy bias (top) and stellar mass distribution (bottom) of the mock LAE catalogs (color points/lines) compared with that derived from observations in the literature (black points/histogram).
    }%
    \vspace*{0.8em} 
\end{figure}

We measure galaxy bias of the mocks by calculating the galaxy two-point correlation function and comparing it to that of the underlying DM at the same epoch. Multiple estimators \citep{1974ApJS...28...19P,1982MNRAS.201..867H,1983ApJ...267..465D,1993ApJ...417...19H,1993ApJ...412...64L} are used; all give consistent results because of the large number ($\sim5\times 10^5$) of LAEs per mock catalog. We Fourier transform the matter power spectrum to obtain the matter two-point correlation function, where the power spectrum is calculated using the Cosmic Linear Anisotropy Solving System (CLASS) package \citep{2011arXiv1104.2932L,2011JCAP...07..034B,2011arXiv1104.2934L}. The linear galaxy bias is obtained using the standard definition
\begin{equation}
b^2 (r) = \frac{\xi_{\rm gal}(r)}{\xi_{\rm m}(r)}
\end{equation}
at $r=8$ Mpc h$^{-1}$ comoving, where the $\xi_{\rm gal}$ and $\xi_{\rm m}$ are the two-point correlation function of galaxies and matter, respectively. The galaxy bias of the set of our four mocks span a range from 1.8--2.4 (Table 2 and the top panel of Figure 1). This agrees well with that of the observed LAEs at roughly the same epoch \citep{2007ApJ...671..278G,2008ApJS..176..301O,2010ApJ...723..869O,2010ApJ...714..255G}, where a small $\lesssim5\%$ fraction of LAEs with X-ray detection have been excluded from these observational clustering analyses. X-ray AGN hosts are found to be more clustered \citep{2011ApJ...736...99A,2014ApJ...796....4A}, and thus the inclusion of these objects as in this work, though subdominant, should elevate the sample averaged galaxy bias slightly.

The stellar mass distributions of our mock LAE catalogs are shown in the bottom panel of Figure 1 and summarized in Table 2. The hatched histogram indicates that of the observed non-AGN sample of HPS LAEs ($z = 2.87^{+0.40}_{-0.58}$), showing a median ${\rm log} \, (M_* / M_{\odot})$ and 16/84 percentile scatter of $8.74^{+0.60}_{-0.72}$ \citep[converted to Chabrier IMF adopted here;][]{2014ApJ...786...59H}. The decline at both the low and high mass ends of the observed sample are physical: the low-end tail is caused by the declining SFR, thus is the intrinsic Ly$\alpha$ production of low mass galaxies; the high-end tail originates from an increasing dust content of high-mass star-forming galaxies. Incompleteness near the detection limit in $\La^{\rm obs}$ does not propagate to bias the stellar mass distribution because of the intrinsically poor correlation between $\La^{\rm obs}$ and $M_*$. Our mock LAE catalogs show similar $M_*$ distributions with that of the observed LAEs, particularly in the low median values of log\,($M_*$) and a similar wide spread, which is about twice large as that of the (more massive) Lyman break galaxies in the same epoch \citep[e.g.,][]{2007ApJ...670..156D}. LAEs in general represent a heterogeneous population of objects with various levels of gravitational clustering, manifested in a large cosmic variance of the galaxy bias.

Using the empirical modeling of Ly$\alpha$ production and dust attenuation, our mock LAEs successfully reproduce the observed galaxy bias and stellar mass distribution simultaneously. For both these properties, the Mock I appears to best match the observed star-forming LAEs. With a small fraction of AGN hosts included, we consider the Mock II ($b=2.0$) as the fiducial mock of the observed galaxy tracers. With this set of mocks, we will discuss the fate of the large-scale structure at $z=2.44$ in HPS-COSMOS, and its uncertainty given the uncertainty in the clustering properties of the galaxy tracers.

\section{Large-scale Structure at $z=2.44$}
Here we present the large-scale galaxy concentration found at $z=2.44$ using the sample of LAEs in HPS-COSMOS supplemented by continuum-selected galaxies with photo-$z$ in COSMOS/UltraVISTA. The field of view of the HPS-COSMOS is of the same order as the characteristic angular size of proto-clusters predicted \citep{2013ApJ...779..127C}. However, the survey probes an order of magnitude longer depth along the line of sight.

\subsection{Redshift Distribution}

In the 71.6 arcmin$^2$ field of view of the HPS-COSMOS (outlined in Figure 3), the redshift distributions of LAEs, photo-$z$ selected galaxies, and the stellar mass volume density of the photo-$z$ galaxies smoothed to a large super-halo scale all show a significant peak at $z\sim2.44$ (Figure 2).

\begin{figure}[!]
     \begin{center}
     \label{redshift_distribution}
        \hspace*{-0.2em}
        \subfigure{%
           \includegraphics[width=0.48\textwidth]{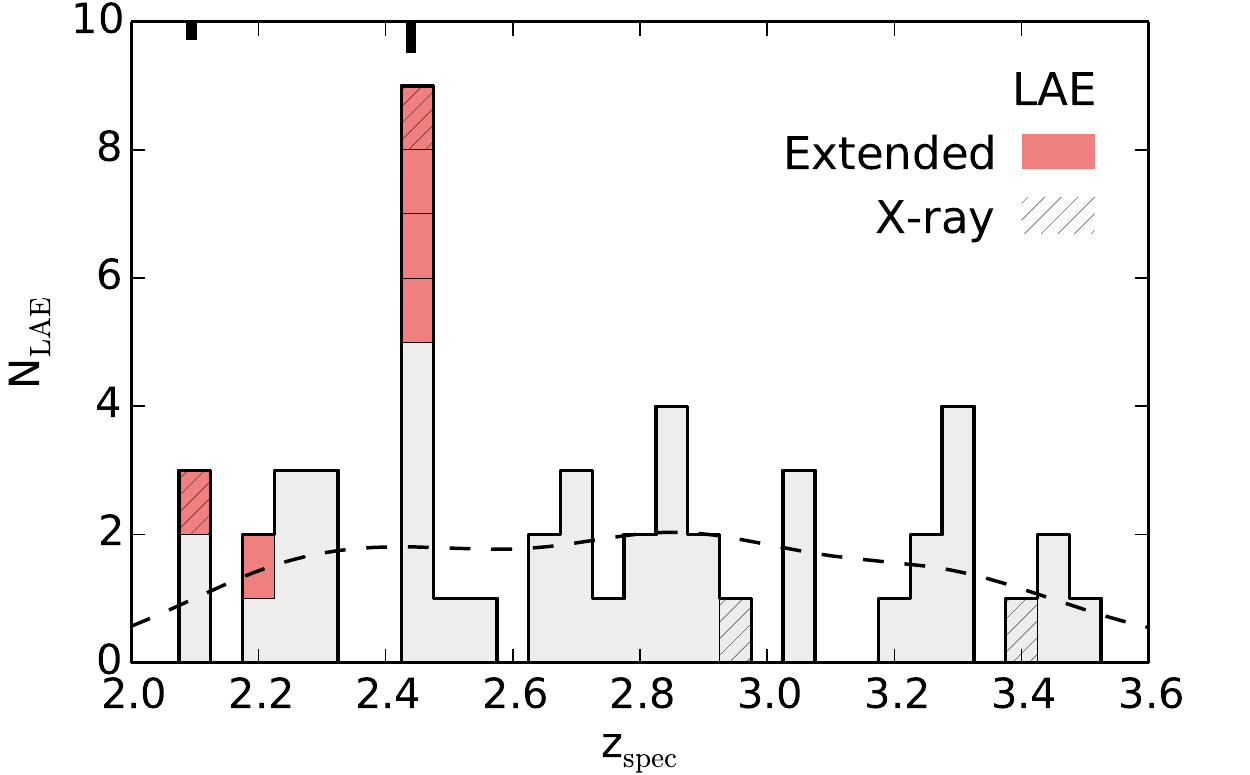}
        }\\ 
        \vspace*{-0.6em}
        \hspace*{-0.2em}
        \subfigure{%
           \includegraphics[width=0.48\textwidth]{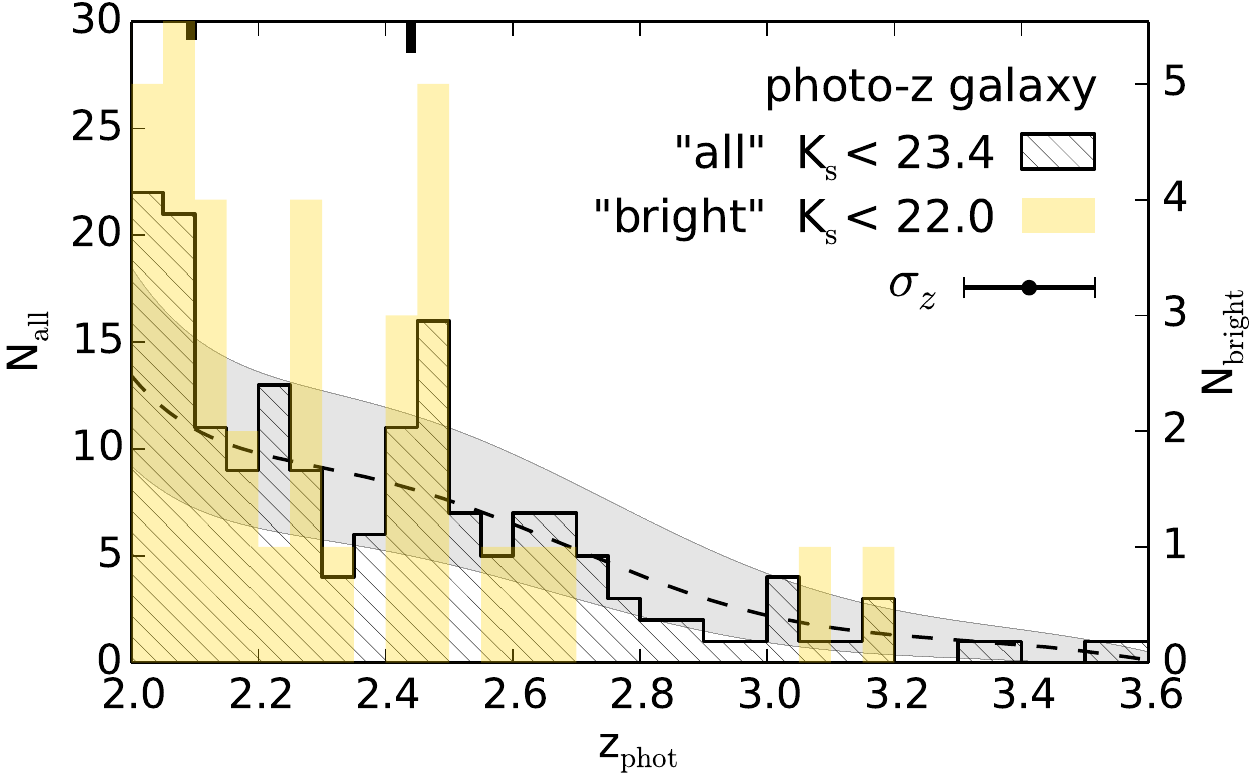}
        }\\ 
        \vspace*{-0.6em}
        \hspace*{-0.2em}
        \subfigure{%
           \includegraphics[width=0.48\textwidth]{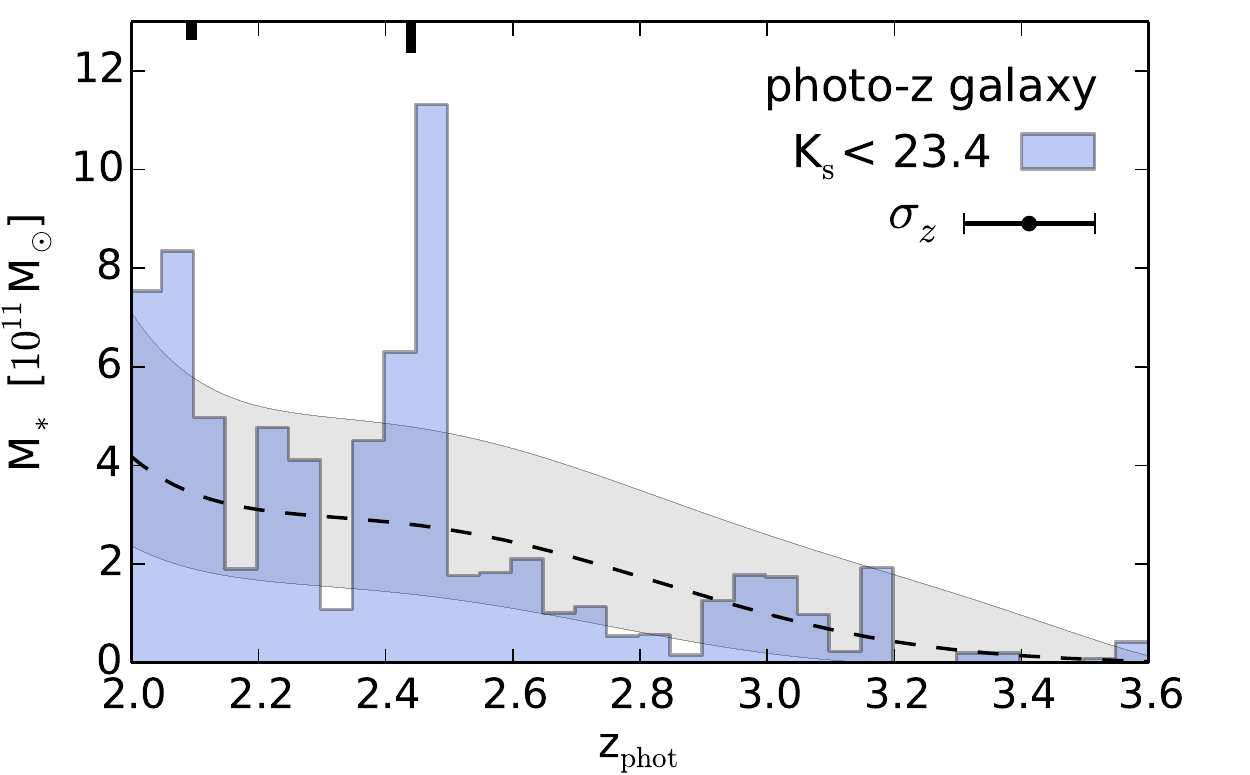}
        }\\ 
        \vspace*{-0.8em}
    \end{center}
    \caption{%
    The redshift distributions of LAE number count (top), continuum-selected galaxy number count with photometric redshifts (middle), and the volume combined stellar mass derived from SED fittings of the continuum-selected galaxies (bottom) in the 71.6 arcmin$^2$ HPS-COSMOS field. The typical photometric redshift error of individual continuum-selected galaxies is shown in error bars. Each redshift bin of a width of 0.05 corresponds to a comoving volume of $\sim 8.5\times12.0\times43.5$ $\rm{Mpc^3\ h^{-3}}$ (at z=2.5). Dashed lines indicate the ensemble averages per redshift bin for each quantity, respectively. The gray regions in the middle and bottom panels show the 68\% scatter per redshift bin for each quantity (only scatter for K$_s<23.4$ is shown in the middle panel) estimated by randomly sampling the whole $\sim1.6$ deg$^2$ COSMOS field. The long and short thick ticks indicate the redshifts of the HPS proto-cluster and a proto-cluster found in the Z-FOURGE survey \citep[][see the Appendix]{2012ApJ...748L..21S}, respectively.
    }%
    \vspace*{0.8em} 
\end{figure}

The top panel of Figure 2 shows the line of sight distribution of 51 LAEs from $z=2.0$ to 3.6 in HPS-COSMOS. The dashed line shows the ensemble average redshift distribution derived from the whole sample of LAEs in 4 HPS fields, smoothed with a Gaussian kernel of $\sigma=0.15$ in redshift (normalized to indicate the expected number of LAEs per redshift bin of 0.05 in the field of view of HPS-COSMOS). A concentration of nine LAEs in the bin at $z=2.45$ is clearly seen. Their mean Ly$\alpha$ redshift is 2.441, indicated by a long thick tick. The ensemble average number density of LAEs at this redshift is $4.0 \times 10^{-4}$ $\rm{Mpc^{-3}\ h^{3}}$. Within the redshift-space bin corresponding to a comoving volume of $8.5\times12.0\times43.5$ $\rm{Mpc^3\ h^{-3}}$, the ensemble average LAE number $\langle \rm N_{LAE} \rangle$ is 1.8. The LAE galaxy overdensity, 
\begin{equation}
\rm{\delta_{LAE}\equiv \frac{N_{LAE}-\langle N_{LAE} \rangle}{\langle N_{LAE} \rangle}},
\end{equation}
is $\sim 4$, averaged over this redshift-space bin.\footnote{The $\rm \delta_{LAE}$ is scale dependent, thus needs to be interpreted carefully.}

The density peak is unlikely to arise from a Poisson sampling of a spatially homogeneous density field, with a $p$-value of $2\times10^{-5}$. Although it is well known that galaxies are clustered, it suggests that the peak is a genuine large-scale structure of physical origin instead of a statistical fluctuation. The value of $\rm \delta_{LAE}$ together with the moderately low LAE bias of $\sim2$ suggest a matter overdensity of $\sim2$, implying that even at this large scale, the matter density field has already evolved to the nonlinear regime. Based on both the linear theory of spherical collapse \citep[e.g.,][]{1999coph.book.....P} and the observational signatures of cluster progenitors expected in $\rm{\Lambda}$CDM cosmological simulations \citep{2013ApJ...779..127C}, the overdensity of this structure at $z=2.44$ appears more than sufficient for it to collapse and evolve into a cluster ($>10^{14}$ $M_{\odot}$) by $z=0$. In \S\,4 we will study the fate of the overdensity in more detail by comparing the observed LAE distribution with the mock LAE catalogs described above.

The middle panel of Figure 2 displays the photo-$z$ distribution of continuum-selected galaxies in the field of HPS-COSMOS (approximated by an $8.46'\times8.46'$ square region). The ``bright'' galaxy sample with $K_s < 22.0$ is shown by the yellow histogram (right $y$-axis), and the whole $K_s < 23.4$ sample is represented by the black hatched histogram. The typical photo-$z$ error of $\sigma_z=0.03\,(1+z)$ at $z=2.5$ is indicated in the figure legend. The dashed line and gray shaded region are the median and 16/84 percentile scatter of the number counts of $K_s < 23.4$ galaxies as a function of redshift, calculated by randomly sampling the whole COSMOS field. Although not shown here, the median redshift distribution for the bright sample of $K_s < 22.0$ would differ slightly, and the scatter would be larger than the gray region for $K_s < 23.4$ galaxies due to both a larger shot noise and a higher cosmic variance (higher intrinsic clustering). Both the $K_s < 22.0$ and $K_s < 23.4$ galaxy number counts clearly reveal a density peak at $2.4<z_{\rm phot}<2.5$ coinciding with the highest LAE concentration in HPS. The overdensity appears to be more pronounced for bright/massive galaxies, which has been previously seen in other massive proto-clusters \citep{2005ApJ...626...44S}. \cite{2014ApJ...782L...3C} compared the $z\sim2.45$ density peak traced by the identical sample of $K_s < 23.4$ galaxies with a large set of matched SAM lightcones (post-processed with observational selection effects and redshift errors), and found that even under this level of redshift uncertainties, the overdensity in photo-$z$ galaxies suggests, with a $\sim 70\%$ confidence level, that this structure will evolve to a cluster with $M_{vir}>10^{14}$ $M_{\odot}$ by $z=0$. We will see in \S\,4 that the LAE distribution with precise redshifts provides consistent but much stronger constraints on the fate of the structure.

The bottom panel of Figure 2 shows, with the blue histogram, the photo-$z$ distribution of stellar mass combining all the continuum-selected galaxies ($K_s < 23.4$) within each redshift-space bin. The dashed line and gray region show the median and 16/84 percentile scatter of this distribution estimated by randomly sampling the whole COSMOS field. Similar to the previous case of galaxy number count, photo-$z$ errors largely smooth out the fluctuation, and slightly reduce the (apparent) cosmic variance, which dominates the gray region. A peak at $2.4<z_{\rm phot}<2.5$ is, again, clearly present. A stellar mass overdensity $\delta_{*}$ can be defined as
\begin{equation}
\delta_*\equiv \frac{\rho_*-\langle \rho_* \rangle}{\langle \rho_* \rangle},
\end{equation}
where $\rho_*$ and $\langle \rho_* \rangle$ are the stellar mass density calculated in a given window and the cosmic stellar mass density at the same epoch, respectively. The $\delta_{*}$ of the most significant bin at $2.45<z_{\rm phot}<2.5$ is $\sim3$, with a signal-to-noise ratio of $\gtrsim4$, which is much higher than that of the number counts of the same galaxy sample shown previously. This difference is related to the fact that there is a higher fractional excess of bright galaxies in this structure as shown previously. The inclusion of faint galaxies also plays a role in reducing the noise. The scatter shown with the gray region includes not only the cosmic variance but also the shot noise of galaxy counts and the systematics in SED fitting.

\begin{figure*}[t!]
     \begin{center}
     \label{sky_map}
        \hspace*{-0.8em}
        \subfigure{%
           \includegraphics[width=0.675\textwidth]{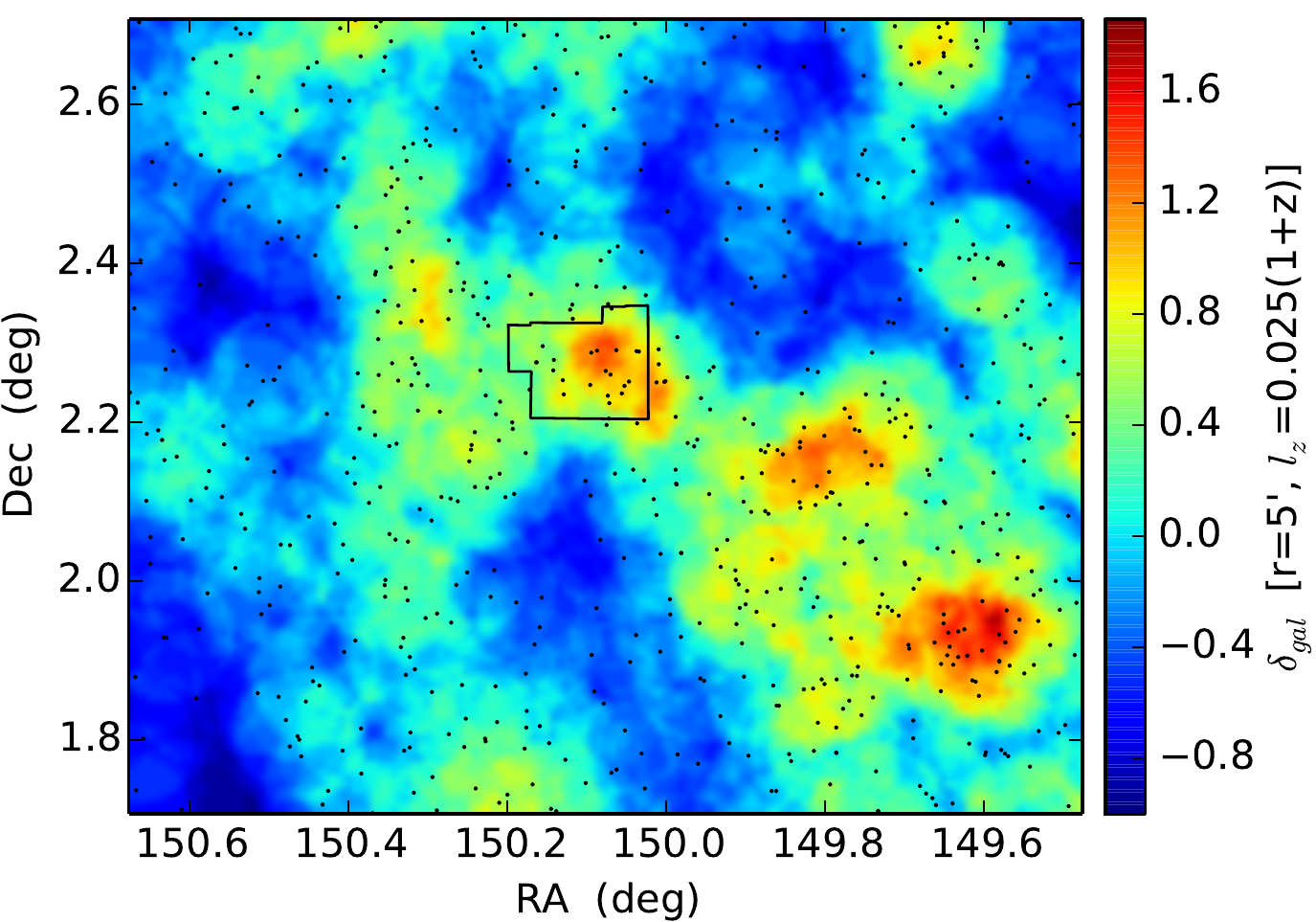}
        }\\ 
        \vspace*{-0.0em}
        \hspace*{-0.8em}
        \subfigure{%
           \includegraphics[width=0.675\textwidth]{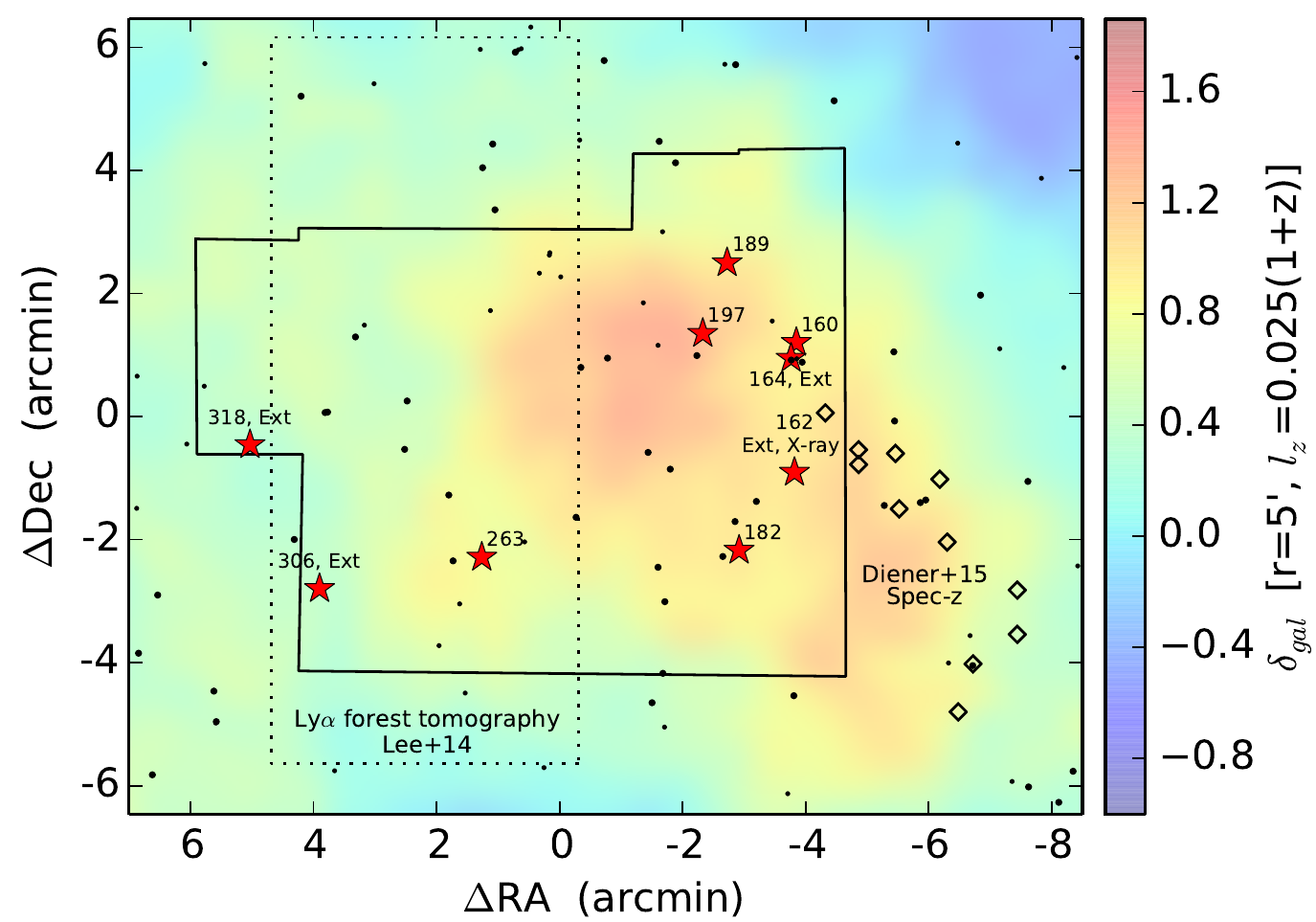}
        }\\ 
        \vspace*{-0.8em}
    \end{center}
    \caption{%
    Sky map of the galaxy distribution at $z\sim2.44$ for the $1.2 \times 1.0$ deg$^2$ COSMOS (top) and the zoomed in of the HPS-COSMOS field indicated by the black outline (bottom). The background color map in both panels shows the density of continuum-selected galaxies with photo-$z$ ($K_s < 23.4$) smoothed with a cylindrical window of $r=5$ and a depth $l_z$ of $\sigma_z = 0.025(1+z)$ as presented in \cite{2014ApJ...782L...3C}. Dots in the top panel represent the galaxy sample used to calculate the large-scale density map within a photo-$z$ full width of $\sigma_z$, and additional ones within a photo-$z$ full width of $2 \sigma_z$ are marked in the bottom panel with smaller symbols. In the bottom panel, stars indicates HPS LAEs. The diamonds indicate continuum-selected LBGs with spectroscopic redshift confirmed in the zCOSMOS survey and the observations in \cite{2015ApJ...802...31D}. The dotted outline represents the Ly$\alpha$ forest tomography field observed by \cite{2014ApJ...795L..12L}. 
            }%
    \vspace*{0.8em} 
\end{figure*}

\subsection{Projected Spatial Distribution}

The $z=2.44$ structure can be seen in the distribution of photo-$z$ galaxies projected on the sky. The top panel of Figure 3 presents the overdensity map of continuum-selected galaxies in the central $1.2 \times 1.0$ deg$^2$ of COSMOS in a thin redshift slice centered at $z_{\rm phot}=2.45$. Dots represent galaxies with a $z_{\rm phot}$ within a full width of $\sigma_z$. This map was generated (but not shown) in the work of \cite{2014ApJ...782L...3C} to search for cluster progenitors. We have smoothed the galaxy distribution with a scale of $\sim 15$ Mpc comoving that corresponds to the typical angular size of proto-clusters. Galaxy overdensity, $\delta_{gal}$ (as defined in Equation (4)), is calculated in a cylindrical window with a radius $r = 5'$ and a redshift depth full width of $l_z = \sigma_z = 0.025\,(1+z)$. Regions of local $\delta_{gal}$ maxima were then identified, and compared with that in a set of matched SAM galaxy lightcones. The three overdense regions shown in red are strong candidate proto-clusters of $M_{z=0} > 10^{14}$ $M_{\odot}$, with a confidence level of $\sim 70\%$. The feature close to the field center corresponds to the HPS-COSMOS $z=2.44$ structure discussed in this work, where the HPS field is outlined in black.\footnote{The other two, at least equally prominent photo-$z$ overdensities in this map have their $\delta_{gal}$ peak in redshift slices near but not in this slice, which correspond to candidate proto-clusters PC17 ($z=2.42$) and PC20 ($z=2.48$), respectively, in \cite{2014ApJ...782L...3C}. They are potentially more massive structures, but the uncertainties in mass overdensity are much larger than that of the HPS $z=2.44$ structure with LAE redshifts presented in this paper. The confirmations of these two structures require spectroscopic follow-ups.} This overdensity roughly fills the whole field of HPS-COSMOS and extends few arcmins to the west. The size of this structure is on the order of 20 Mpc comoving, consistent with that of massive cluster progenitors studied in simulations \citep{2006ApJ...646L...5S,2013ApJ...779..127C,2014arXiv1412.1507S}. 

In the bottom panel of Figure 3 we expand the scale to show the HPS-COSMOS field. The background $\delta_{gal}$ map is the same as that shown in the top panel. The nine LAEs in the redshift spike presented in \S\,3.1 are indicated by red stars. Dots represent continuum-selected galaxies with a photometric redshift within $2.45 \pm \sigma_z$, and those within $2.45 \pm 0.5 \sigma_z$ are marked by larger symbols.

\subsection{Stellar Mass of the Continuum-selected Galaxies}

Continuum-selected galaxies ($K_s < 23.4$) inside the HPS-COSMOS field with $2.35<z_{\rm phot}<2.5$ have a median stellar mass of $4.5^{+8.1}_{-3.5} \times 10^{10}$ $M_{\odot}$ (among a sample of 33), which is about double of that of $2.1^{+5.0}_{-1.3} \times 10^{10}$ $M_{\odot}$ for galaxies outside the overdensity with the same $K_s$-band limit and redshift.

\subsection{Substructures}

\begin{figure}[t!]
     \begin{center}
     \label{substructure}
        \hspace*{-0.2em}
        \subfigure{%
           \includegraphics[width=0.48\textwidth]{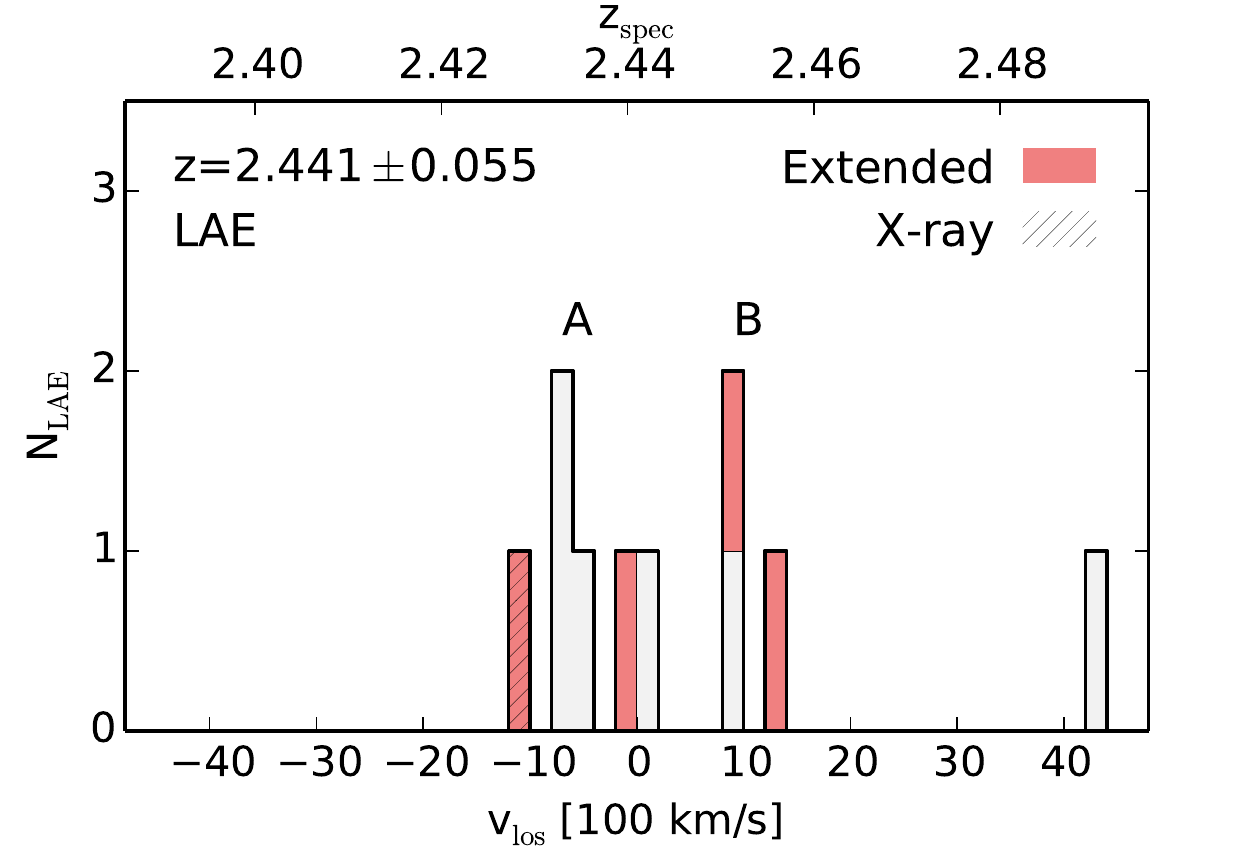}
        }\\ 
       \vspace*{-0.8em}
    \end{center}
    \caption{%
    The line of sight velocity $v_{\rm los}$ distribution of HPS-COSMOS LAEs centered at $z=2.441$. Red and hatched elements indicate LAEs with an extended Ly$\alpha$ halo and X-ray counterpart, respectively.
     }%
     \vspace*{0.8em} 
\end{figure}

Figure 4 shows the detailed line of sight velocity $v_{\rm los}$ distribution of HPS-COSMOS LAEs centered at $z=2.441$ (the mean redshift of the nine LAEs in the overdensity). The nine LAEs span a full range of $\sim 2500$ km s$^{-1}$ in $v_{\rm los}$, with a dispersion $\sigma_{\rm v,\, los}$ of 905 km s$^{-1}$ \citep[using the gapper estimator for small $N$ in][]{1990AJ....100...32B}. Based on the large spatial extent of the structure on the projected sky, this high $\sigma_{\rm v,\, los}$ is unlikely to be dominated by peculiar velocities of a collapsed structure. There appear to be two substructures labeled A and B in Figure 4 (hereafter groups A and B, though the term ``group'' here does not refer to galaxies in a common parent halo). These substructures show a $\sigma_{\rm v,\, los}$ of 456 and 221 km s$^{-1}$ for groups A and B, respectively, with a separation of $\sim 1600$ km s$^{-1}$ in their mean velocities. This separation corresponds to a line of sight comoving distance of $22.4$ Mpc, which is larger than the HPS-COSMOS field size of $14.5$ Mpc on the sky. Indeed, group A and B both have their members scattered across the entire HPS-COSMOS field on the projected sky.

\subsection{Other Evidence of the Structure in the Literature}
Using spectroscopic redshifts of continuum-selected galaxies in the zCOSMOS-deep survey, \cite{2013ApJ...765..109D} identified 42 ``proto-groups'' in COSMOS at $1.8<z<3.0$. These systems were identified using a working definition of associations of $\ge 3$ galaxies that pass a linking length criterion,\footnote{The algorithm in \cite{2013ApJ...765..109D} is designed to identify groups or group progenitors, thus capturing overdensities with a scale smaller than that considered in this work for cluster progenitors. Their galaxy selection based on broad-band colors and limiting magnitudes ($K_s < 23.5$; $B < 25.3$) typically excludes LAEs, which by definition, have a large excess of Ly$\alpha$ with respect to the stellar continuum.} and are expected to each assemble into a single halo by $z=0$. Strikingly, the richest structure \citep[five galaxies, ID 22 in][]{2013ApJ...765..109D} found in this large volume is located immediately west of the HPS structure at the same redshift of $2.44$. It also coincides with the spatial extent of photo-$z$ galaxy overdensity as shown previously in \S\,3.2 and Figure 3. \cite{2015ApJ...802...31D} spectroscopically confirmed a total of 11 galaxies (diamonds in the bottom panel of Figure 3) and gave a central redshift of 2.45. They suggest that this structure will collapse to form a massive cluster of $10^{14} - 10^{15}$ $M_{\odot}$ by $z=0$. With their spectroscopic campaign in a wider field, this result strongly suggests that the HPS $z=2.44$ structure is indeed large and associated with an extremely rare density concentration.

\cite{2014ApJ...795L..12L} presented a 3-dimensional cosmic density reconstruction in a $5' \times 11.8'$ field in COSMOS at $2.20 \le z \le 2.45$ using tomography of Ly$\alpha$ absorption seen in the spectra of bright background galaxies. This field (dotted line in the bottom panel of Figure 3) coincides with the east half of the HPS-COSMOS field. As shown in the Figure 3 of \cite{2014ApJ...795L..12L}, there is a strong and complex overdensity of Ly$\alpha$ absorbing gas (the densest among the survey volume) at $2.43 \lesssim z \lesssim 2.45$, coinciding with our HPS LAE overdensity at $z=2.44$. Their figure shows another three spectroscopically-confirmed, broad-band selected LBGs \citep[from][]{2007ApJS..172...70L,2015A&A...576A..79L} in this structure. This result independently supports the large-scale structure seen in HPS-COSMOS at $z=2.44$.

\def\arraystretch{1.5}
\begin{centering}
\begin{deluxetable*}{lccccccccc}
\tablecaption{Properties of Mock Structures of LAEs at $z=2.4$} \tablewidth{0pt}
\tablehead{	\colhead{ID}	&
		\colhead{${\rm log} \ M_{z=0}$$\rm$\tablenotemark{a}}	&
		\colhead{$\rm N_{\rm merged}$$\rm$\tablenotemark{b}}	&
		\colhead{$\Delta \theta_{\rm pc}$$\rm$\tablenotemark{c}}	&
		\colhead{$\rm D_{\rm los,\, app}$$\rm$\tablenotemark{d}}	&
		\colhead{$\rm D_{\rm los,\, int}$$\rm$\tablenotemark{e}}	&
		\colhead{$\sigma_{\rm v,\, los,\, app}$$\rm$\tablenotemark{f}}	&
		\colhead{$\sigma_{\rm v,\, los,\, int}$$\rm$\tablenotemark{g}}	&
		\colhead{$\sigma_{\rm v,\, los,\, app}^{\, A}$$\rm$\tablenotemark{h}}	&
		\colhead{$\sigma_{\rm v,\, los,\, int}^{\, A}$$\rm$\tablenotemark{i}} \\
		\colhead{}	&
		\colhead{$[M_{\odot}]$}	&
		\colhead{}	&
		\colhead{[arcmin]}	&
		\colhead{[Mpc]}	&
		\colhead{[Mpc]}	&
		\colhead{[km s$^{-1}$]}	&
		\colhead{[km s$^{-1}$]}	&
		\colhead{[km s$^{-1}$]}	&
		\colhead{[km s$^{-1}$]}
		}
\startdata
Mock I & $14.58_{-0.53}^{+0.38}$	&	$3.24 \pm 1.99$	&	$5.69_{-2.77}^{+5.21}$	&	$31.62_{-4.83}^{+4.78}$	&	$37.42_{-5.98}^{+5.56}$	&	$764_{-147}^{+128}$	&	$239_{-94}^{+76}$	&	$304_{-134}^{+151}$	&	$221_{-92}^{+104}$ \\
\bf{Mock II} & $\bf{14.49_{-0.35}^{+0.45}}$	&	$\bf{3.23 \pm 1.92}$	&	$\bf{6.38_{-3.13}^{+4.24}}$	&	$\bf{32.43_{-5.36}^{+6.04}}$	&	$\bf{37.69_{-5.56}^{+6.90}}$	&	$\bf{757_{-136}^{+144}}$	&	$\bf{229_{-83}^{+93}}$	&	$\bf{285_{-128}^{+166}}$	&	$\bf{202_{-85}^{+115}}$ \\
Mock III & $14.53_{-0.46}^{+0.41}$	&	$3.49 \pm 1.96$	&	$5.66_{-2.87}^{+4.68}$	&	$32.63_{-6.66}^{+5.57}$	&	$37.74_{-6.29}^{+8.21}$	&	$773_{-156}^{+143}$	&	$231_{-87}^{+109}$	&	$280_{-126}^{+172}$	&	$204_{-96}^{+128}$ \\
Mock IV & $14.52_{-0.35}^{+0.39}$	&	$3.71 \pm 1.99$	&	$5.67_{-2.70}^{+4.13}$	&	$32.06_{-5.27}^{+5.77}$	&	$37.40_{-6.12}^{+6.69}$	&	$748_{-152}^{+151}$	&	$230_{-83}^{+94}$	&	$278_{-142}^{+159}$	&	$210_{-98}^{+114}$
\enddata
\tablenotetext{a}{Median Virial mass of the most massive $z=0$ descendant DM halo (friends-of-friends group central) of a mock structure.}
\tablenotetext{b}{Number of LAEs in each mock structure that will be merged into the same friend of friend group by $z=0$.}
\tablenotetext{c}{Angular separation between the field center targeting a mock structure and the true center of the corresponding proto-cluster (defined to be the center of mass of its member DM halos).}
\tablenotetext{d}{Full size (comoving) of a mock structure of nine LAEs along the line of sight in redshift space.}
\tablenotetext{e}{Full size (comoving) of a mock structure of nine LAEs along the line of sight in real space.}
\tablenotetext{f}{Line of sight velocity dispersion of a mock structure in redshift space.}
\tablenotetext{g}{Line of sight velocity dispersion of a mock structure in real space (peculiar velocity only).}
\tablenotetext{h}{Line of sight velocity dispersion of the main substructure (criterion 2 in the text) in redshift space.}
\tablenotetext{i}{Line of sight velocity dispersion of the main substructure (criterion 2 in the text) in real space (peculiar velocity only).}
\end{deluxetable*}
\end{centering}

\section{Cosmic Evolution of the Structure}

We now examine the fate of the HPS-COSMOS large-scale-structure at $z=2.44$ using the mock LAE catalogs constructed in \S\,2.3. A large number of realizations of simulated HPS-COSMOS observations are generated. First, for each simulation box (500 Mpc h$^{-1}$ comoving) of the four mock LAE catalogs, we generate three projected pseudo-lightcones from the $z=2.4$ snapshot with a viewing angle along the $x$, $y$, and $z$ axes, respectively. Specifically, the apparent redshift of each LAE is determined by its line of sight position (for the component of the Hubble expansion) and peculiar velocity. Galaxy properties are non-evolving to focus on the comparison at $z\sim2.44$. Second, we target each pseudo-lightcone with a large number of fields of $8.46'\times8.46'$ that match the area of HPS-COSMOS, each probing a pencil-beam like volume. Third, regions similar to the observed $z=2.44$ overdensity are identified as mock structures. The main constraints provided by the observations are the level of LAE overdensity and their distribution along the line of sight, including the substructures described above. We define a set of criteria to identify mock structures in simulations: (1) there must be nine LAEs within a full span of 26.7 to 47.2 Mpc comoving along the line of sight, which correspond to the 2-$\sigma$ limits of that observed for the HPS-COSMOS overdensity,\footnote{Here the number count of LAEs is considered definite, as the effects of the shot noise on the proto-cluster characterization will be captured automatically by selecting a large realizations of mock structures. The uncertainty in the full span of the structure is estimated by bootstrapping the structure centered distances of the nine observed LAEs, with an additional contribution from instrument error ($\sigma_{\rm v,\, los} = 130$ km s$^{-1}$).} and (2) to match the more compact substructure of group A, six out of the nine LAEs are required to be in a line of sight interval within 23.6 Mpc comoving, the 1$\sigma$ upper limit of that observed. These criteria select a few thousand mock structures per mock LAE catalog. A fraction of the structures represent the same underlying structures seen under different viewing angles and/or covered by different realizations of the HPS-COSMOS pointing on the sky (i.e., different field centers). Finally, we examine the relation between these high-redshift mock LAE structures and their $z=0$ descendant halos.

Table 3 summarizes the main properties of mock HPS structures at $z=2.4$ and their descendants at $z=0$. This structure has a large line of sight extent in redshift-space of $\rm D_{\rm los,\, app} \sim 32$ Mpc comoving. Excluding the contribution from peculiar velocities, the simulations show that its real space full size $\rm D_{\rm los,\, int}$ is $\sim38$ Mpc, larger than its $\rm D_{\rm los,\, app}$. This is a classic signature of the Kaiser effect \citep{1987MNRAS.227....1K}, suggesting that the outermost shell (as in the picture of spherical collapse scenario) has already decoupled from the cosmic expansion and started to collapse in comoving space. Most of the LAEs in the structure occupy a distinct DM halo during the observed epoch; these halos have already been influenced by self-gravity as an ensemble, and will combine to form larger halos in later epochs. The most massive $z=0$ descendant halo of this structure is expected to have a virial mass of $10^{14.5\pm0.4}$ $M_{\odot}$ ($\sim 90\%$ probability with $M_{z=0} > 10^{14}$ $M_{\odot}$), corresponding to a massive galaxy cluster. Only $\gtrsim3$ LAEs in the structure will be merged onto this main halo by $z=0$, and are considered to be the true members of the proto-cluster. It is less certain whether the secondary substructure, group B, can evolve to a cluster-scale halo at $z=0$. Since the whole structure at $z=2.44$ has already broken away from the Hubble flow, we expect a gravitationally bound, but not entirely virialized, descendant structure at $z=0$ (with a size of several physical Mpc) containing a massive cluster.

Although the structure is most likely to be a genuine proto-cluster with $M_{z=0} > 10^{14}$ $M_{\odot}$, there is a $\sim10\%$ chance that the most massive $z=0$ descendant halo will have a smaller virial mass of $10^{13.5}$--$10^{14}$ $M_{\odot}$. In this case the structure would be considered as a massive proto-group \citep[e.g.,][]{2013ApJ...765..109D}. Such a slightly lower mass overdensity is often associated with cosmic web filaments, which have been studied in more details at lower redshifts \citep{2013ApJ...779..139S,2014ApJ...796...51D,2014MNRAS.439.2571H,2015arXiv150206602S}.

The inferences of the mass overdensity and $z=0$ virial mass would stay the same if we exclude X-ray detected LAEs and trace the structure using star-forming LAEs only. In this case the $z=2.44$ overdensity consists of eight LAEs instead of nine, while a lower biased mock galaxy population (Mock I with $b=1.82$) would be considered as fiducial to interpret the observation, resulting in a nearly identical level of inferred mass overdensity. We caution that our results would be biased if the Ly$\alpha$ escape fraction were to depend strongly on large-scale environment. However, a strong environment effect would result in a galaxy two-point correlation function that significantly departs from the power-law form measured for typical star-forming galaxies and DM halos in simulations on relevant scales. Such a departure is not seen for observed LAEs \citep{2007ApJ...671..278G,2007ApJ...668...15K,2008ApJS..176..301O,2010ApJ...723..869O,2010ApJ...714..255G,2015arXiv150101215B}.

\section{Extended L\MakeLowercase{y}$\alpha$ halos and AGNs}

In the top panel of Figure 2 and in Figure 4, we label the extended Ly$\alpha$ sources in red. As described in \S\,2.1, these systems are robustly ruled out to be point sources, with diameters of several tens of physical kpc (see the the Ly$\alpha$ surface brightness profiles of the most extended sources in \cite{2011ApJS..192....5A}). Strikingly, an enhancement of extended LAEs in large-scale overdensities is present. Five out of six extended LAEs in HPS-COSMOS are in large-scale overdense regions: four in our HPS-COSMOS structure at $z=2.44$ and another one in a $z=2.10$ structure discovered in the ZFOURGE survey \citep{2012ApJ...748L..21S,2014ApJ...795L..20Y} with three LAEs detected in HPS-COSMOS (see the Appendix). The tendency for extended LAEs to be in overdense regions is highly significant against random fluctuations with a $p$-value of $3 \times 10^{-4}$. Within the $z=2.44$ structure, four out of the total nine LAEs are extended. These four extended LAEs are distributed in both group A and group B (Figure 4), making this environment---Ly$\alpha$ blobs correlation prominent at a scale at least equal or larger than a cluster progenitor, as we have shown that the $z=0$ descendants of the whole $z=2.44$ structure will still be collapsing around a massive virialized cluster. Also, HPS-261 (see Table 4 in the appendix), the extended LAE associated with the ZFOURGE $z=2.10$ structure, is $\sim 2.5$ Mpc (physical) away from the well-confirmed density peaks of the proto-cluster \citep{2014ApJ...795L..20Y}. Therefore, the properties of circumgalactic-scale Ly$\alpha$ emission appear to be directly or indirectly connected to the elevated DM and baryon density on a super-halo scale.

Four LAEs are detected in X-ray (hatched regions in the top panel of Figure 2 and Figure 4). Their X-ray luminosity ($L_X \sim 10^{44}-10^{45}$\,erg\,s$^{-1}$), being $\sim 3-20$ times larger than their observed Ly$\alpha$ luminosity, implies that AGN photoionization likely dominates the intrinsic Ly$\alpha$ production in these systems. Two out of the four AGNs are in known large-scale overdensities: one in the z=2.10 structure (see the Appendix) and one in the z=2.44 structure. This result gives a moderately low $p$-value of 0.2 against the null hypothesis that AGNs are a random subset of the bulk LAE population drawn from an uniform probability distribution. 

Two out of the total six extended Ly$\alpha$ sources in HPS-COSMOS are associated with X-ray detected AGNs. The scenario that extended Ly$\alpha$ halos tend to host AGNs is significant against random fluctuation, with a $p$-value of 0.06. Furthermore, these AGN powered Ly$\alpha$ halos are all in overdense regions, implying possible causations behind the correlation of environment, AGNs, and the size of Ly$\alpha$ emission.

\begin{figure*}[!]
    \begin{center}
    \label{redshift_distribution}
        \hspace*{-0.6em}
        \subfigure{%
            \includegraphics[width=0.48\textwidth]{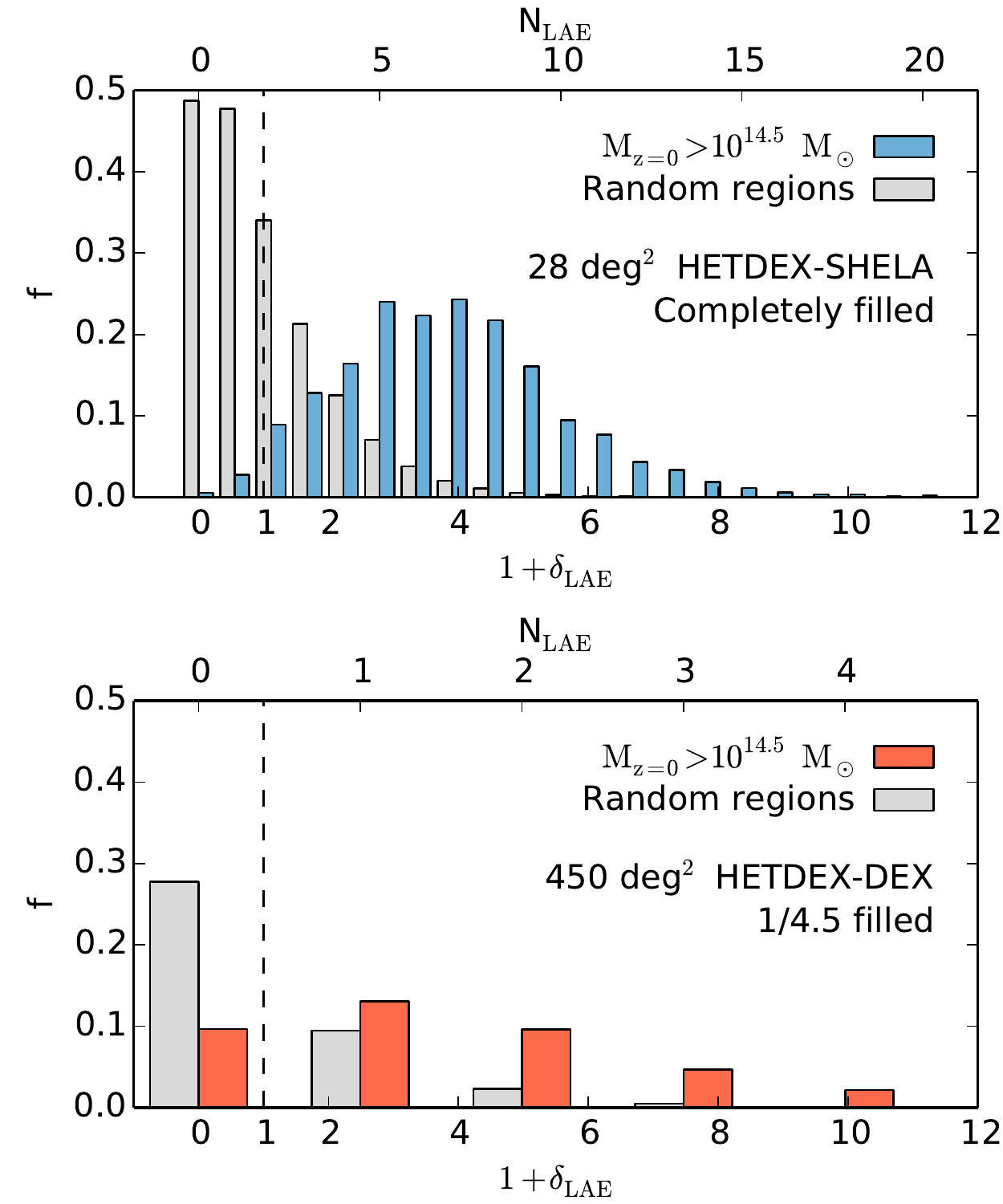}
        }%
        \hspace*{1.2em}
        \subfigure{%
           \includegraphics[width=0.48\textwidth]{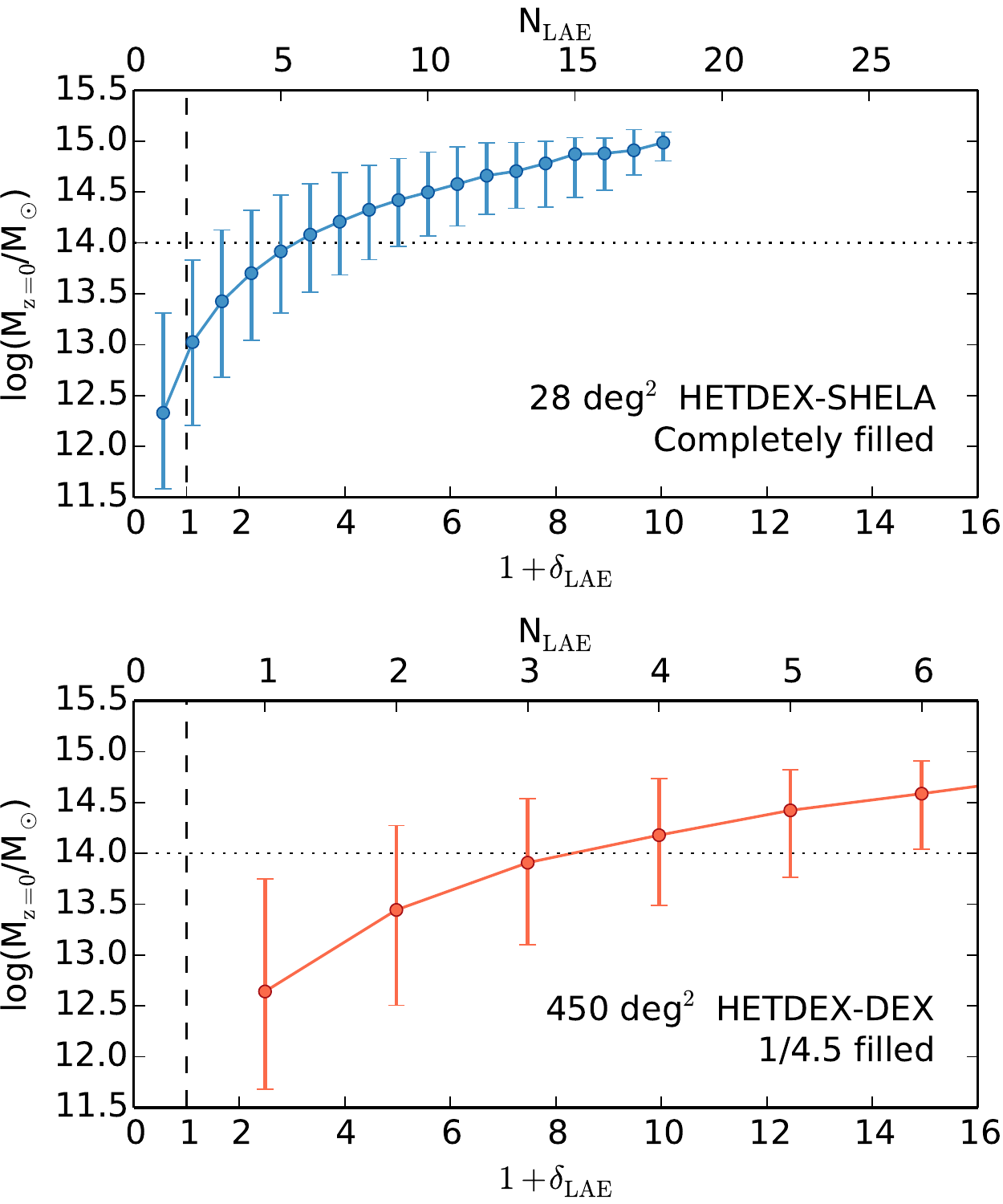}
        }\\ 
        \vspace*{-0.8em}
    \end{center}
    \caption{%
    \textbf{Left panels:} the probability distribution of LAE overdensity $\rm \del_{LAE}$ globally (gray histograms) and regions centered on proto-clusters with $M_{z=0} > 10^{14.5}$ $M_{\odot}$ (color histograms) in simulations for the 28 deg$^2$ HETDEX-SHELA (top) and the $\sim 450$ deg$^2$ dark energy survey of HETDEX (bottom) where 1/4.5 of the area will be covered by IFU fibers. The $\rm \del_{LAE}$ is measured in a cylindrical window of $r=6$ Mpc\,h$^{-1}$ comoving and $l_{los} = 20$ Mpc\,h$^{-1}$ comoving. \textbf{Right panels:} median and 16/84 percentile scatter of the $z=0$ descendant halo mass $M_{z=0}$ as a function of $\rm \del_{LAE}$ for each HETDEX baseline fields, evaluated by sampling the whole volume of the simulations randomly.
    }
    \vspace*{0.8em} 
\end{figure*}

\section{Proto-clusters in the HETDEX Survey}

Using the HET and VIRUS, the coming HETDEX survey will perform blind spectroscopy of order a million LAEs at $1.9<z<3.5$, allowing construction of cosmic density maps for galaxy environmental studies and select a large and homogeneous sample of cluster progenitors. To precisely measure the matter power spectrum at the peak scale of BAO for dark energy science, the main survey (300 deg$^2$ Spring field plus 150 deg$^2$ Fall field, hereafter, HETDEX-DEX) will sample the large area sparsely \citep{2013JCAP...12..030C}, with a $1/4.5$ spatial filling factor (the fraction of sky area covered by IFU fibers). This would impact, unfortunately, the performance of localized studies in real space through increasing shot noise. However, in a $28$ deg$^2$ area within the Fall field overlapping with the $Spitzer$-HETDEX Exploratory Large Area (SHELA; PI: Papovich) survey and other ancillary photometry (hereafter HETDEX-SHELA); complete coverage (unity filling factor) will be achieved by multiple dithering. Here we examine the performance of proto-cluster identification expected in HETDEX-DEX and HETDEX-SHELA with a counts in cell algorithm applied to our mock LAE catalogs. This analysis essentially uses the correlation between high-redshift local LAE overdensity $\rm \del_{LAE}$ and the $z=0$ descendant halo mass $M_{z=0}$ under the inclusion of observational effects and realistic noise. Implicitly, the input cosmology, gravitational structure formation, and galaxy formation model in the simulation together are used as the prior of the analysis. The difference in the survey filling factor of our two baseline fields here allows us to demonstrate the effects of a generic noise source in density mapping---the shot noise that arises from a discrete and finite sampling of the underlying parent distribution.

Under the wavelength-dependent line sensitivity of HETDEX and assuming a Ly$\alpha$ luminosity function of \cite{2007ApJ...667...79G} for LAEs with no redshift evolution between $1.9<z<3.5$, the expected comoving number density of HETDEX LAEs is nearly flat at $\sim 8 \times 10^{-4}$ Mpc$^{-3}$ at $1.9<z<2.5$, twice that as in HPS, and decreases to $\sim 3 \times 10^{-4}$ Mpc$^{-3}$ at $z=3.5$. For HETDEX-DEX (1/4.5 filled) and HETDEX-SHELA (completely filled), we generate a mock LAE catalog at $z=2.4$, based on the LAE modeling described in \S\,2.3. These two catalogs have the same clustering properties as the Mock II used for characterizing the HPS $z=2.44$ structure.\footnote{The slightly deeper Ly$\alpha$ luminosity limit of HETDEX compared with that of HPS is expected to have only a limited effect on the bias of LAEs \citep{2008MNRAS.391.1589O}.} They have different ensemble average LAE number densities, $\rm n_{\,HETDEX-SHELA}:n_{\,HPS}:n_{\,HETDEX-DEX}=1:1/2:1/4.5$. We implement this feature by tuning for each mock the survival probability described in \S\,2.3 to account for the uncertain stochasticity of Ly$\alpha$ escape, plus, for the case of the HETDEX-DEX, the incompleteness due to a sub-unity survey filling factor.

We then perform a counts in cell analysis of LAE overdensity in the mocks and examine its dependency on the $z=0$ descendant halo mass. A redshift-space cylindrical window of $r=6$ Mpc\,h$^{-1}$ comoving and $l_{los} = 20$ Mpc\,h$^{-1}$ comoving (including peculiar velocity) is used to calculate local LAE number $\rm N_{LAE}$ and overdensity $\rm \del_{LAE}$. This window is ideal for the observed density contrast at $z\gtrsim 2$ between proto-clusters of $M_{z=0} \sim 10^{14.5}$ $M_{\odot}$ and field, while a more sophisticated optimization can be performed by varying the window with redshift, targeting $M_{z=0}$, and the filling factor. Thus the performance of proto-cluster identification presented below should be viewed as a lower limit.

The left panels of Figure 5 show, at $z=2.4$, the expected probability distribution of $\rm \del_{LAE}$ globally (gray histograms) and that of the regions centered on proto-clusters with $M_{z=0} > 10^{14.5}$ $M_{\odot}$ (color histograms) in the simulation. The upper and bottom panels show the expected results for the HETDEX-SHELA and HETDEX-DEX surveys, respectively. These $\rm \del_{LAE}$ distributions represent a same intrinsic correlation between large-scale mass budget and their $z=0$ collapsed mass modulated by different levels of shot noise, which fractionally scales with roughly the inverse square root of the true population mean number per window (characterized approximately by a Poisson process). In the case of HETDEX-SHELA, proto-clusters show a significantly higher $\rm \del_{LAE}$ compared to the ensemble, where a threshold in $\rm \del_{LAE}$ can be used to separate proto-cluster regions from field. In the case of the HETDEX-DEX, only proto-clusters with the highest $\rm \del_{LAE}$ can be separated, thus producing a much lower completeness.

Since non-proto-cluster regions occupy the bulk of cosmic volume and can appear dense due to sampling noise and the intrinsic scatter (usually subdominant), it needs to be quantified how well the $M_{z=0}$ can be recovered given a measured $\rm \del_{LAE}$. We show this correlation for each HETDEX baseline field in the right panels of Figure 5. $M_{z=0}$ is the virial mass of the most massive $z=0$ descendant halo (friends-of-friends group central) of the LAEs within a window for measuring $\rm \del_{LAE}$. The dots and errorbars indicate, respectively, the median and 16/84 percentile scatter of the $M_{z=0}$ at a given $\rm \del_{LAE}$. For HETDEX-SHELA, the $M_{z=0}$--$\rm \del_{LAE}$ correlation is fairly tight. The scatter in $M_{z=0}$ shrinks from $\sim 1.5$ dex at $\rm \del_{LAE} \sim 0$ to $\lesssim 0.5$ dex at $\rm \del_{LAE} \sim 10$, showing that the most massive proto-clusters, while being rare, can be identified robustly in HETDEX-SHELA. On the other hand, the larger shot noise (horizontal scatter in nature) in HETDEX-DEX not only extends the range of possible $\rm \del_{LAE}$ (also shown in the left panels) but also increases the scatter of this correlation. The median $M_{z=0}$-$\rm \del_{LAE}$ correlation in HETDEX-DEX lies everywhere below that in HETDEX-SHELA. This result is due to the upward scatter from intrinsically less dense regions, which outweighs the downward scatter because of the much higher abundance of the former, thus the estimated $M_{z=0}$ for a genuine dense structure is biased low when the noise is finite. If the structure shows other evidence of overdensity like in the case of the HPS structure at $z=2.44$, a deeper Ly$\alpha$ observational program is likely to increase the best estimated $M_{z=0}$, and asymptotically approach the true value when having a large $N$. 

For $M_{z=0} > 10^{14}$ $M_{\odot}$ proto-clusters and a required purity of 70\%, 80\%, 90\%, the completeness in HETDEX-SHELA is $\sim50\%$, 30\%, 15\%, respectively; in the case of HETDEX-DEX, the completeness decreases to 5\%, 1\%, and nearly 0\%, respectively, as the lower scatter in the bottom-right panel of Figure 5 never reaches much above $10^{14}$ $M_{\odot}$. These estimates represent the minimum performance. An ideal strategy for the case of the wide HETDEX-DEX would be focusing on finding the largest and rarest proto-clusters, where an even larger window can be beneficial since these structures remain overdense on a large scale. A large window is also preferred for a statistical reason---the shot noise, which roughly scales with the volume of the window to the $-$3/2 power, can be reduced. Unfortunately in this case the accuracy of the positional centering and the handle of substructure remain poor. Additional investigations of these densest structures in HETDEX-DEX are needed to calculate their exact overdensity, and would supplement a massive sample to that found in HETDEX-SHELA.

Conservatively, we expect to obtain a sample (>90\% confidence) of a few tens of $M_{z=0} > 10^{15}$ $M_{\odot}$ proto-clusters and a few hundreds of $M_{z=0} > 10^{14.5}$ $M_{\odot}$ in HETDEX-SHELA at $1.9<z<3.5$; and another hundred $M_{z=0} \sim 10^{15}$ $M_{\odot}$ proto-clusters in HETDEX-DEX.

\section{Discussion}

Here we focus our discussion on proto-cluster identification quantified in terms of $M_{z=0}$, the comparison between the HPS structure and other known high-redshift overdensities in the literature, and the dependency of galaxy properties on large-scale environment.

\begin{figure*}[t!]
     \begin{center}
     \label{comparison}
        \hspace*{0em}
        \subfigure{%
           \includegraphics[width=0.48\textwidth]{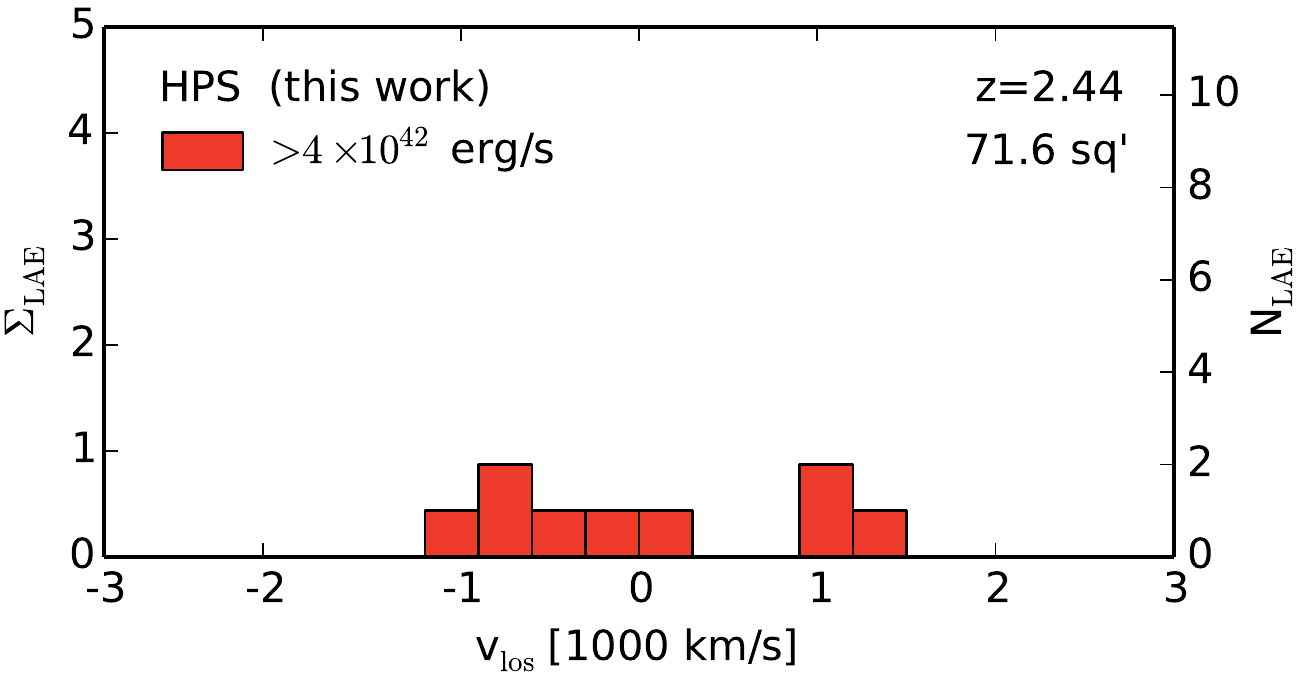}
        }\\ 
        \vspace*{0em}
        \hspace*{0em}
        \subfigure{%
           \includegraphics[width=0.48\textwidth]{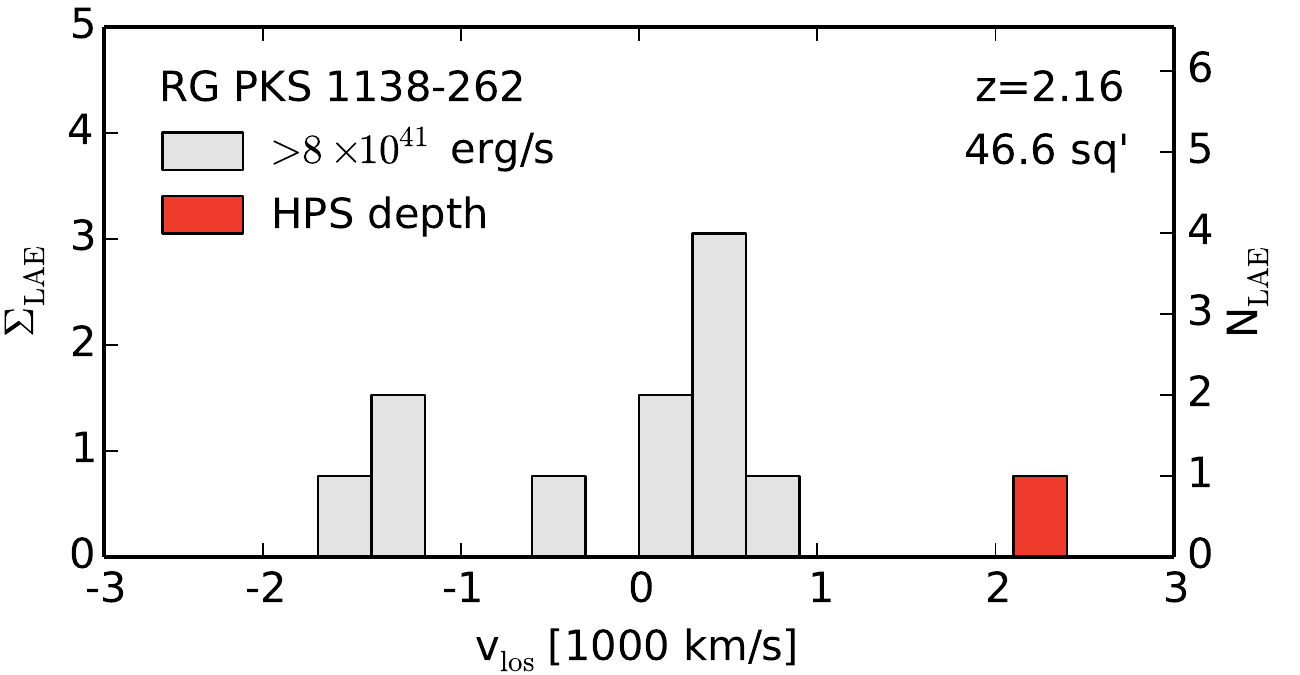}
           \hspace*{0.5em}
           \includegraphics[width=0.48\textwidth]{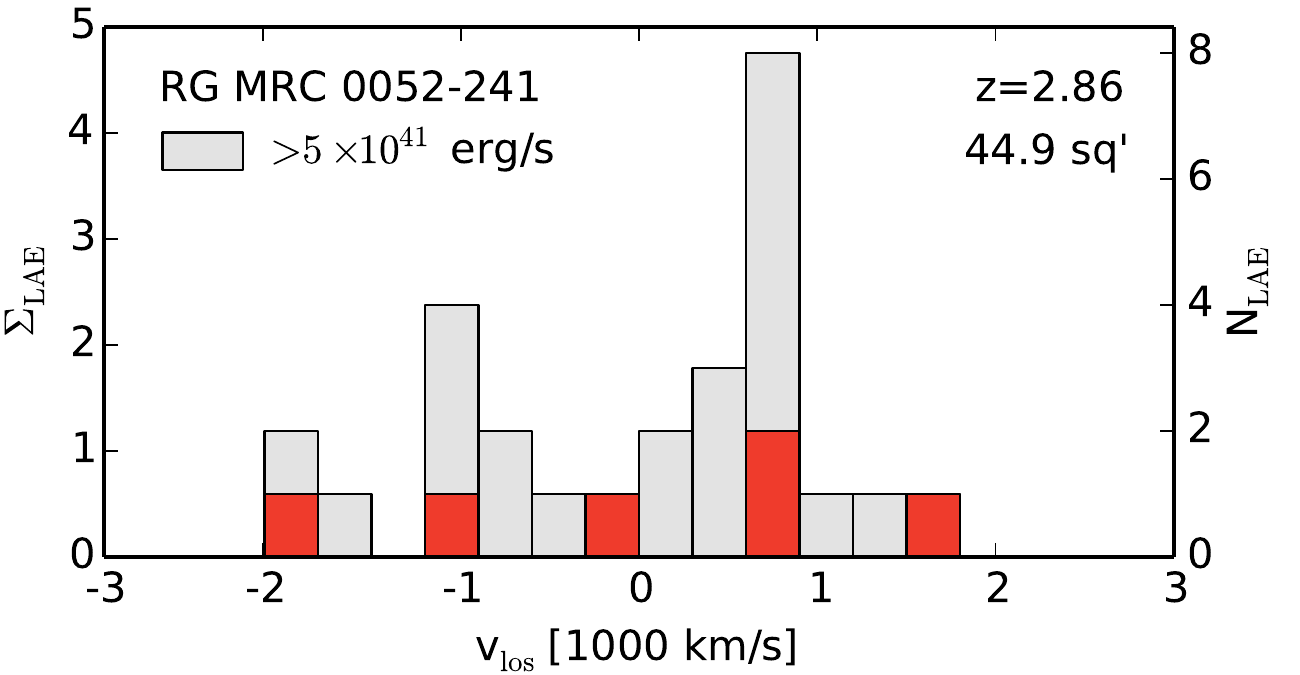}
        }\\ 
        \vspace*{0em}
        \hspace*{0em}
        \subfigure{%
           \includegraphics[width=0.48\textwidth]{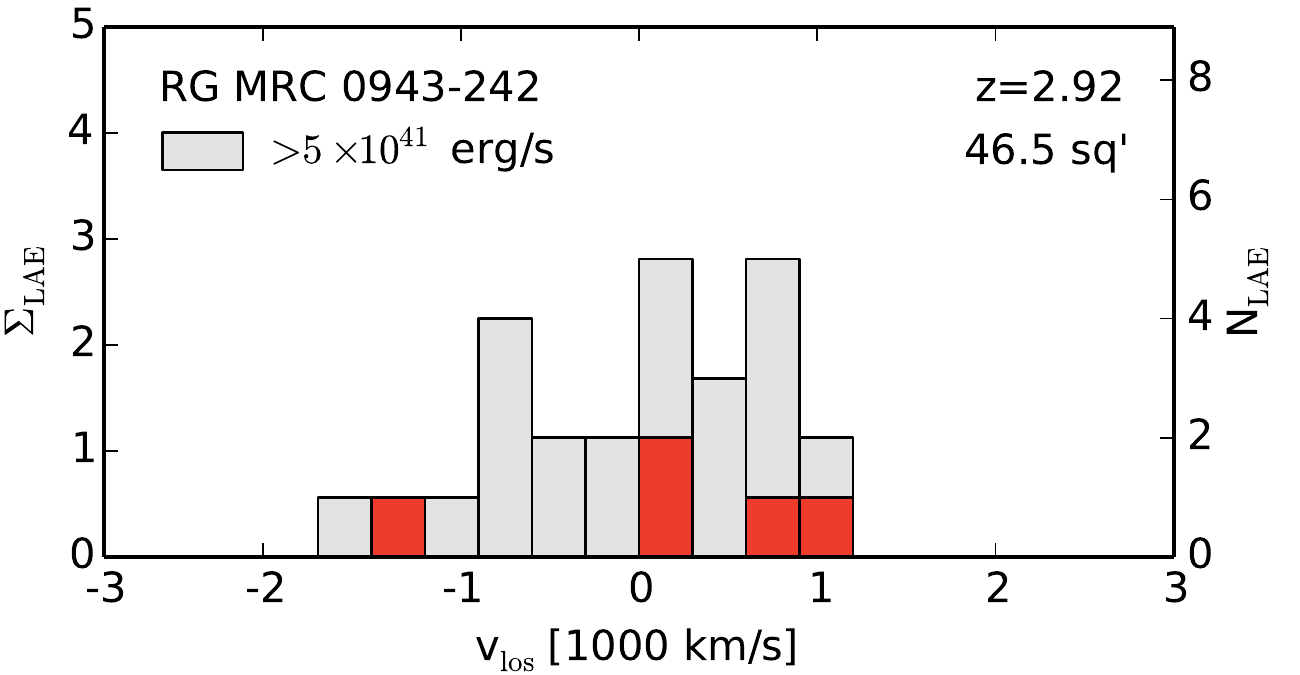}
           \hspace*{0.5em}
           \includegraphics[width=0.48\textwidth]{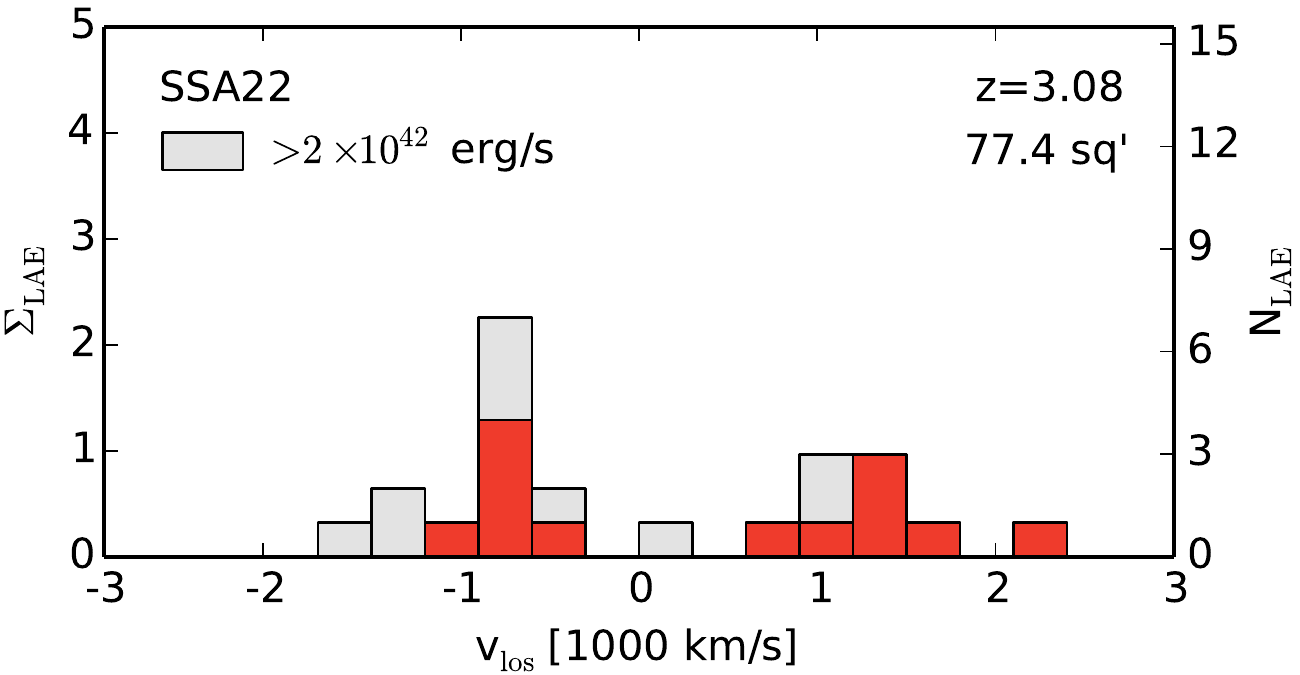}
        }\\ 
        \vspace*{0em}
        \hspace*{0em}
        \subfigure{%
           \includegraphics[width=0.48\textwidth]{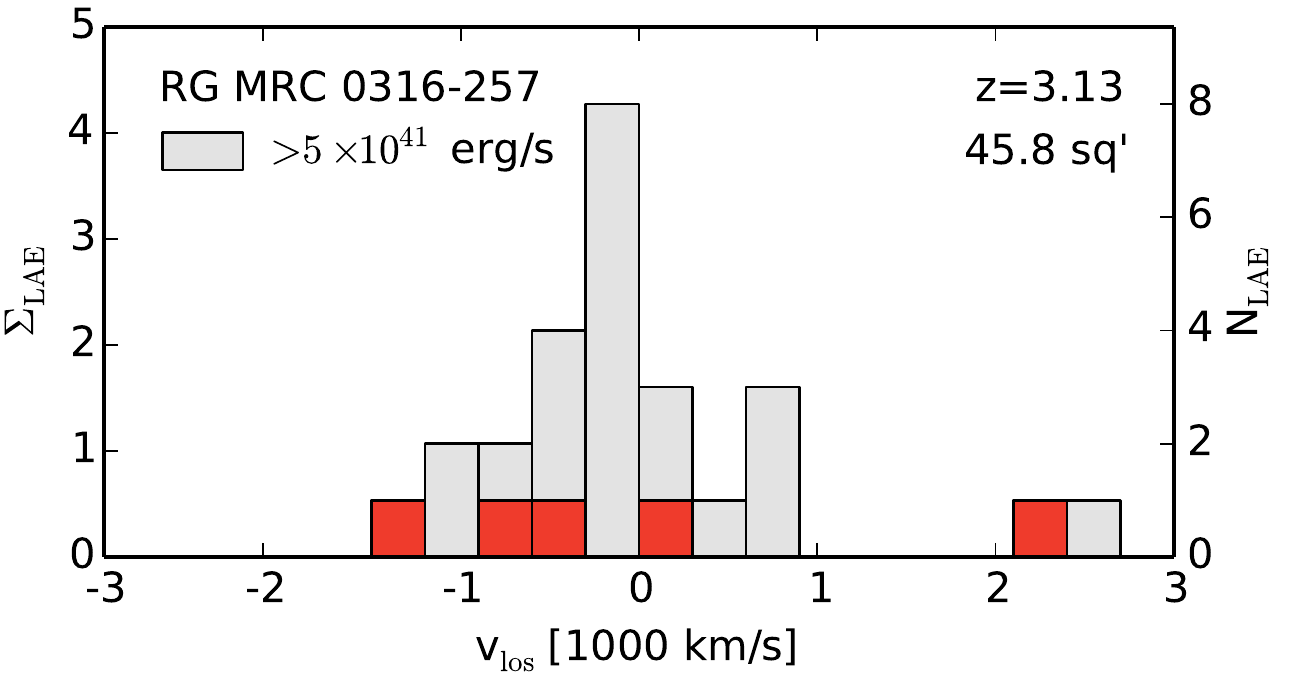}
           \hspace*{0.5em}
           \includegraphics[width=0.48\textwidth]{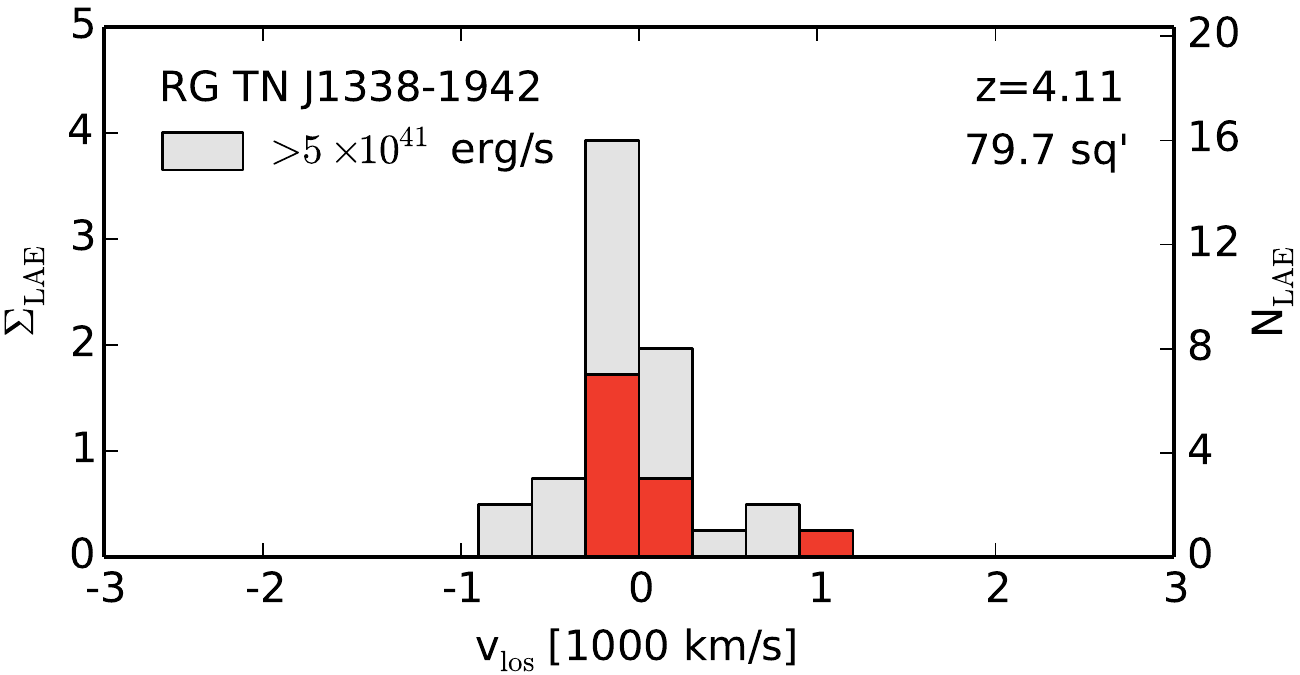}
        }\\ 
        \vspace*{0em}
    \end{center}
    \caption{%
    Line of sight velocity distribution of the LAEs in the HPS $z=2.44$ structure compared with that of proto-clusters in the literature around powerful radio galaxies \citep[RG;][]{2000A&A...358L...1K,2004A&A...428..793K,2000A&A...361L..25P,2002ApJ...569L..11V,2005A&A...431..793V,2007A&A...461..823V} and the SSA22 field \citep{1998ApJ...492..428S,2000ApJ...532..170S,2004AJ....128.2073H,2005ApJ...634L.125M,2012ApJ...751...29Y}. The histograms are normalized to the same scale of surface number density in an arbitrary unit of number per comoving area (left $y$-axis). Red histograms represent LAEs above the Ly$\alpha$ luminosity limit of HPS, and gray histograms show, in the case of radio galaxy proto-clusters, LAEs with deeper Ly$\alpha$ luminosity limits indicated in the figure legend. Slightly different EW$\rm_{Ly\alpha}$ criteria of $>20$ \AA\ and $>15$ \AA\ were adopted in the selections of LAEs in HPS and RG fields, respectively; while a stricter criterion of EW$\rm_{Ly\alpha} > 40$ \AA\ was adopted in the case of SSA22.
            }%
    \vspace*{0.8em} 
\end{figure*}

In \S\,4, we identified regions that match the HPS $z=2.44$ structure in the four mock LAE catalogs (\S\,2.3) of different clustering within the uncertainty of that observed. These four mocks essentially yield the same prediction on the $z=0$ descendant cluster mass of $10^{14.5\pm0.4}$ $M_{\odot}$. A key reason for this result is that the abundance of $z=0$ clusters, quantified by the $z=0$ halo mass function, posts a strong prior in determining the fate of the observed high-$z$ overdensity (especially in the massive end). A higher LAE overdensity than that of the HPS $z=2.44$ structure would not increase the inferred $M_{z=0}$ substantially; instead, it might pose challenges to the concordance cosmology (in particular, $\Omega_{m}$ and $\sigma_{8}$) as the probability of finding such a density peak will be extremely low. For Mock I to IV, only 3\% to 9\% of the HPS-COSMOS realizations (of the same survey volume) have a region that meets the mock structure criteria. Therefore our discovery of this dense $z=2.44$ structure is perhaps a great coincidence. However, the cluster interpretation is consistent with the ``proto-group'' study in the zCOSMOS-deep survey \citep{2013ApJ...765..109D} that no larger structure is found as traced by their spectroscopic sample in the field several times larger than that of the HPS-COSMOS. Our results also imply that an LAE bias of much lower than 2 (thus a higher inferred mass density) would produce a conflict with the concordance cosmology. Similar problems would be raised if the clustering of LAEs was not modeled (see \S\,2.3) to have a realistically large cosmic variance as constrained by their wide stellar mass distribution.

In \S\,3.1, we showed that the HPS $z=2.44$ structure has a higher signal-to-noise ratio in stellar mass overdensity than in galaxy number counts using the same set of continuum-selected galaxies with photo-$z$. In \S\,3.3, we demonstrated that the stellar mass of galaxies in the overdensity is about twice that outside the overdensity. These results suggest that the onset of star formation in this structure occurred at significantly earlier epochs, supporting the picture of the ``cosmic downsizing.'' The result agrees quantitatively well with that found in a proto-cluster in the quasar HS1700+643 field at $z=2.30$ \citep{2005ApJ...626...44S}, and is consistent with the high formation redshifts inferred from stellar population synthesis of the low-redshift cluster red sequence \citep[e.g.,][]{2010ApJ...709..512R}. It remains to be tested, with a future large sample of proto-clusters, whether the stellar mass excess is generic for high-redshift overdensities.

The cumulative star formation (stellar mass) in a region is a direct consequence of the past accretion and cooling of baryons triggered by the gravitational field of the total matter enclosed, whereas the density contrast in terms of pure number counts in a dense region can be reduced by galaxy mergers as structure/galaxy formation progresses. Therefore it is expected that the stellar mass density field traces the underlying matter density field more tightly. We suggest that in the case of photometric surveys with or without subsequent spectroscopy (where stellar mass can be better measured than in emission-line galaxy surveys), an analysis of stellar mass density contrast should ideally replace galaxy number counts as a standard technique to (1) define galaxy environment and identify possible environmental effects, (2) recover the underlying matter field, and (3) identify proto-clusters and predict their $M_{z=0}$. For most photometric surveys, the resources required for measuring stellar mass do not exceed that for measuring photometric redshift to a sufficient accuracy. Thus a boost of performance for the aforementioned applications can be expected, at no extra cost.

The HPS $z=2.44$ structure does not have a significantly high level of total instantaneous SFR estimated by SED fitting of the continuum-selected galaxies. This result is in line with the general understanding that galaxy star formation is considerably bursty and could be triggered by sporadic and instantaneous accretion of cold streams from the cosmic web \citep{2009Natur.457..451D} or violent disk instabilities \citep{2009ApJ...703..785D,2009ApJ...706..203O}. Therefore a measure of the large-scale SFR density field would be quite noisy compared to that of the stellar mass.

We compare the HPS $z=2.44$ structure studied in this work with other proto-clusters in the literature. Figure 6 shows the line of sight velocity distribution of LAEs in the HPS $z=2.44$ structure (same with Figure 4), five previously known LAE overdensities around powerful radio galaxies \citep{2000A&A...358L...1K,2004A&A...428..793K,2000A&A...361L..25P,2002ApJ...569L..11V,2005A&A...431..793V,2007A&A...461..823V}, and the structure in the SSA22 field at $z=3.08$ \citep{1998ApJ...492..428S,2000ApJ...532..170S,2004AJ....128.2073H,2005ApJ...634L.125M,2012ApJ...751...29Y}. These comparison structures were observed with deep narrow-band imaging at the wavelength of Ly$\alpha$ of the radio galaxies or, in the case of SSA22, of a serendipitously discovered overdensity in a redshift survey of continuum-selected galaxies. A slightly more relaxed EW$\rm_{Ly\alpha}$ criterion of $>15$ \AA\ (compared to the $>20$ \AA\ used in HPS) was adopted in the LAE selection in the radio galaxy fields, while a stricter criterion of EW$\rm_{Ly\alpha} > 40$ \AA\ was used in the case of SSA22. These narrow-band selected LAEs were then investigated by slit spectroscopy, revealing a redshift-space concentration of a few tens of LAEs for each (gray histograms). Similar to the HPS structure, the structures around radio galaxies are expected to each evolve to a galaxy cluster of several times 10$^{14}$ $M_{\odot}$ by $z=0$ based on the level of LAE overdensity observed; the overdensity of the continuum-selected LBGs in the SSA22 field suggests a slightly higher $z=0$ cluster mass of $\sim 10^{15}$ $M_{\odot}$, \citep[see summaries and discussion in][]{1998ApJ...492..428S,2007A&A...461..823V,2013ApJ...779..127C}. The detection limit in terms of the Ly$\alpha$ luminosity for the HPS $z = 2.44$ structure is relatively shallow compared to these comparison proto-cluster fields. Taking this limiting luminosity into account, the HPS structure shows a LAE excess that is comparable to all the comparison structures (except for PKS 1138-262, which lacks very bright LAEs). In fact, the comoving number density of LAEs in the HPS structure is higher than that of all the radio galaxy structures, and similar to that of the SSA22 structure if observed down to the same HPS depth (red histograms).\footnote{To compare observations with different field of view and at different redshifts, we normalize the histograms to the same scale of surface number density in an arbitrary unit of inverse comoving area (left $y$-axis).} Thus a large population of faint LAEs might exist for the HPS $z=2.44$ structure, requiring deeper observations to confirm. Similarly, the ZFOURGE $z=2.10$ proto-cluster (see the Appendix) might also hosts a population of faint LAEs yet to be observed. The HPS $z=2.44$ and radio galaxy structures, all having a similar end point in terms of $z=0$ cluster mass, can provide a rough evolutionary picture of early cluster kinematics across cosmic time. In general, the line of sight velocity dispersion of proto-clusters increases from $\lesssim300$ km s$^{-1}$ for TN J1338-1942 at $z=4.11$ to $\sim900$ km s$^{-1}$ for the three structures at $z\sim2.5$ (PKS 1138-262, MRC 0052-241, and MRC 0943-242). However, this latter velocity dispersion might be too large for the structures to collapse entirely by $z=0$ (see \S\,4), and perhaps by coincidence, these three overdensities around radio galaxies all show a bimodal velocity structure. Conversely, the three higher redshift radio galaxy structures (MRC 0943-242, MRC 0316-257, and TN J1338-1942) show a more clear central concentration in velocity space. The perhaps slightly more massive proto-cluster at $z=3.08$ in the SSA22 field shows a double-peak profile of LAE line of sight velocity distribution, with a combined dispersion of $\sim1000$ km s$^{-1}$. Such a velocity distribution suggests, again, that the large structure in the SSA22 field is unlikely to collapse entirely by $z=0$. A massive descendant cluster of $\sim 10^{15}$ $M_{\odot}$ connected with dense filaments, or a pair of slightly lower mass clusters are expected at $z=0$. In conclusion, this comparison demonstrates that proto-clusters, though they can be characterized with $M_{z=0}$ to first order, show a wide variety of topology in the phase-space mass distribution. A larger sample of proto-clusters with different mass and topology at different redshifts is needed for detailed investigations.

In \S\,5 we demonstrated that there is a boost of extended LAEs and a marginal excess of AGNs in large-scale overdensities. Under the scenario of resonant scattering of Ly$\alpha$ in the circumgalactic medium (CGM), the production of Ly$\alpha$ photons and the phase-space distribution of the CGM around galaxies together determine the spatial profile of Ly$\alpha$ emission \citep{2007ApJ...657L..69L,2009ApJ...696..853L,2009ApJ...704.1640L,2011ApJ...739...62Z,2012MNRAS.424.1672D,2012A&A...546A.111V}. We speculate that the super-halo scale galaxy environment might be connected to the excess of extended Ly$\alpha$ halos through (1) triggering AGNs (thus elevating the production of Ly$\alpha$ photons) for luminous LAEs and (2) altering the CGM profile and/or inflow/outflow structures of all galaxies in general. These speculations are supported by the correlations of extended LAEs and environment in our HPS data, and also various reports in the literature.

First, the excess of AGNs in high-redshift large-scale overdensities is found in other systems \citep{2002A&A...396..109P,2009ApJ...691..687L,2013ApJ...768....1M}, possibly triggered by efficient gas accretion and subsequent funneling induced by frequent galaxy interactions. Second, nearly all the luminous Ly$\alpha$ blobs ($\La^{\rm obs} \ge 10^{43.7}$\,erg\,s$^{-1}$) show signatures of obscured quasars \citep{2013ApJ...771...89O}, and many of them are located in dense environments. Third, while extremely deep narrow-band imaging suggests that halos of scattered Ly$\alpha$ are a generic feature of typical high-redshift star-forming galaxies \citep{2011ApJ...736..160S,2014MNRAS.442..110M}, the scale lengths of the Ly$\alpha$ radial profile are small in the field \citep{2013ApJ...776...75F} and significantly elevated as super-halo scale galaxy densities increase \citep{2012MNRAS.425..878M}. The \cite{2012MNRAS.425..878M} comparison was done while controlling for the UV magnitude, which traces the ionizing photons generated by young stars. This correlation thus needs to be explained beyond the amount of intrinsic Ly$\alpha$ production, possibly through a correlation between environment and the phase-space distribution of CGM.

The density gradient of the CGM of galaxies in dense environments might be flattened, as the baryons follow the two-halo term clustering of DM halos at this scale \citep{2011ApJ...739...62Z}. Furthermore, the CGM dynamics and inflow/outflow structures might be affected, manifested in a shorter fallback time scale of the galactic wind launched by galaxies in dense environments \citep{2008MNRAS.387..577O,2011MNRAS.416.1354D}. This effect is expected to be more prominent for low-mass galaxies, where the local gravitational potential is not deep enough to govern entirely the galaxies' baryonic cycle. It remains to be explored how efficient this mechanism can be on the scales exceeding a single DM halo, and how CGM dynamics is connected to the resonant scattering and escape of Ly$\alpha$. In the 28 deg$^2$ HETDEX-SHELA survey, where a complete coverage will be achieved by dithering, Ly$\alpha$ blobs extended significantly beyond a fiber ($1.5''$ diameter, $\sim12$ physical kpc at $z=2.5$, 1/3 of that of HPS) can be identified. The correlation between diffuse Ly$\alpha$ halos and galaxy environment thus can be quantified with high quality statistics.

In \S\,6 we presented the expected performance of proto-cluster identification in HETDEX. As shown in \S\,3 and \S\,4, the nine LAEs in HPS provide a much more stringent constraint on the $M_{z=0}$ of the $z=2.44$ structure than that of the $\gtrsim 30$ continuum-selected galaxies with photo-$z$ derived from $\sim30$ bands in COSMOS/UltraVISTA. Although the technique of stellar mass overdensity enhances the signal of the underlying matter density field, the photo-$z$ errors of continuum-selected galaxies significantly increase the noise level \citep[see Figure 13 in ][]{2013ApJ...779..127C}. The key advantage of emission-line galaxy surveys in proto-cluster searches is the ability to measure precise galaxy redshifts efficiently, thus reducing the effects of line of sight projection. However, broad-band imaging with a wide range of wavelength coverage is still crucial for galaxy population studies. The 28 deg$^2$ HETDEX-SHELA field will be extremely valuable on this subject for its LAE redshifts and deep optical to far-IR photometry ($u$, $g$, $r$, $i$, $z$, $K$, $Spitzer$-3.6, 4.5\,$\rm \mu m$, Herschel-250, 350, and 500\,$\rm \mu m$).

\section{Conclusion}

Galaxy proto-clusters at $z\gtrsim2$ can be found and confirmed efficiently in large emission-line galaxy surveys. In this paper, we presented the discovery and a detailed characterization of a large-scale structure containing a proto-cluster at $z=2.44$ traced by LAEs in the HETDEX Pilot Survey. The same structure is also seen in continuum-selected photometric redshift catalogs, and appears as a significant overdensity in stellar mass density and gas absorption maps. We constructed a set of mock LAE catalogs matching the clustering properties of the observed LAEs and examined the fate of this HPS structure, and demonstrated the expected performance of proto-cluster identification in the full HETDEX survey, which will confirm a large number of structures similar to the one studied here.

\begin{itemize}[leftmargin=*]
  \item The HPS, which performed a LAE survey of $\sim 7' \times 10'$ in COSMOS at $1.9 < z < 3.8$, discovered a prominent density concentration of nine bright LAEs at $z=2.44$. With the photometric redshift galaxy catalog of COSMOS/UltraVISTA, we demonstrated that this structure is also seen in overdensities of continuum-selected galaxies in both number counts and volume-specific stellar mass. The structure extends $\gtrsim 30$ Mpc comoving along the line of sight with two subgroups of LAEs, and a $\sim 20$ Mpc comoving on the sky revealed by the continuum-selected galaxies. Using the zCOSMOS survey and additional spectroscopy, \cite{2013ApJ...765..109D,2015ApJ...802...31D} identified and confirmed a galaxy overdensity adjacent to the HPS-COSMOS field at $z=2.45$, which appears connected to the HPS structure presented in this paper. We find other independent evidence of this structure in the literature, including an excess of Ly$\alpha$ absorbing gas \citep{2014ApJ...795L..12L}.
   
  \item To compare the HPS structure with simulations of cosmic structure formation, we constructed a set of mock LAE catalogs from the SAM of \cite{2013MNRAS.428.1351G}. The LAEs were modeled based on the Ly$\alpha$ production by young stars and an empirical treatment of the escape of Ly$\alpha$ in dusty ISM. The modeling self-consistently reproduces the observed LAE galaxy bias and stellar mass distribution. Regions in the mocks as dense as the HPS $z=2.44$ structure are then identified and tracked to $z=0$. The HPS structure, although spanning a few tens of Mpc comoving, should have already broken away from the Hubble flow. Part of the structure will collapse to form a galaxy cluster with $10^{14.5\pm0.4}$ $M_{\odot}$ by $z=0$.

  \item Four of the nine LAEs in the HPS structure are significantly extended in Ly$\alpha$ emission, and one of them shows an AGN signature in X-ray (is also an extended Ly$\alpha$ source). We speculate that a super-halo scale dense environment might facilitate AGN activities and alter the CGM profiles around high-redshift star-forming galaxies, boosting the spatial extent of Ly$\alpha$. The median stellar mass of the continuum-selected galaxies in the HPS structure is about twice that of the field counterparts. These results demonstrate an accelerated co-evolution of massive galaxies and their supermassive black holes in overdense environments.

  \item Finally, we predict the performance of proto-cluster identification in the coming HETDEX survey, which will observe of order a million LAEs at $1.9<z<3.5$. In the full $\sim450$ deg$^2$ HETDEX, where 1/4.5 of the sky will be covered by IFU fibers, we expect to confirm at least a hundred massive proto-clusters with $M_{z=0} \sim 10^{15}$ $M_{\odot}$. In the $28$ deg$^2$ HETDEX-SHELA field, where a complete sky coverage will be performed, we expect to obtain a few tens of proto-clusters with $M_{z=0} > 10^{15}$ $M_{\odot}$, and a few hundreds of $M_{z=0} > 10^{14.5}$ $M_{\odot}$. Together with a rich set of ancillary photometry, the HETDEX-SHELA field will provide a powerful data set to study the rapid mass assembly and galaxy growth of present day massive clusters in their formation epoch.
\end{itemize}

\newcommand{\jcap}{JCAP}
\newcommand{\nar}{New Astron. Rev.}

\begin{acknowledgements}
We thank McDonald Observatory and its staff for supporting the development of Mitchell Spectrograph and the observations of the HETDEX Pilot Survey. The construction of the Mitchell Spectrograph was possible thanks to the generous support of the Cynthia \& George Mitchell Foundation. YC acknowledges support from the University of Texas at Austin Graduate School Fellowship. The Millennium Simulation databases used in this paper and the web application providing online access to them were constructed as part of the activities of the German Astrophysical Virtual Observatory (GAVO). The K$_{s}$-selected catalog of the COSMOS/UltraVISTA field used in this paper was constructed by \cite{2013ApJS..206....8M}, which contains PSF-matched photometry in 30 photometric bands covering the wavelength range 0.15$\micron$ $\rightarrow$ 24$\micron$ and includes the available $GALEX$ \citep{2005ApJ...619L...1M}, CFHT/Subaru \citep{2007ApJS..172...99C}, UltraVISTA \citep{2012A&A...544A.156M}, S-COSMOS \citep{2007ApJS..172...86S}, and zCOSMOS \citep{2009ApJS..184..218L} datasets. The Institute for Gravitation and the Cosmos is supported by the Eberly College of Science and the Office of the Senior Vice President for Research at the Pennsylvania State University. SJ acknowledges support from NSF grant AST-1413652 and the National Aeronautics and Space Administration (NASA) JPL SURP Program.

\end{acknowledgements}

\bibliographystyle{apj}
\bibliography{HPS.bib}



\appendix

\section{Another Structure at at z=2.10}

HPS-COSMOS also partially covers another proto-cluster at $z=2.10$ (indicated by a short thick tick in Figure 2) discovered by galaxy overdensities in a deep medium-bands photometric survey of ZFOURGE \citep{2012ApJ...748L..21S}. Three cores of possibly virialized halos of $M_{vir}\gtrsim10^{13}$ $M_{\odot}$ at the observed epoch are identified. In \cite{2014ApJ...782L...3C}, we recovered this structure on a scale of $\sim 15$ Mpc comoving using the same COSMOS/UltraVISTA photometric redshift galaxy catalog used here, and together revealed other 35 candidate proto-clusters in COSMOS field. Recently, \cite{2014ApJ...795L..20Y} has performed a large spectroscopic campaign and confirmed more than 50 objects in this structure, estimating a redshift zero virial mass of $M_{z=0}=10^{14.4\pm0.3}\ M_{\odot}$. Within the region of the three cores, we do not detect any LAE in HPS, but we did find three LAEs (summarized in Table 4, see also Figure 2) associated with this $z=2.10$ proto-cluster several arcmins away from the cores, indicating that the overdensity of this structure indeed has a large spatial extent, consistent with what was reported in \cite{2014ApJ...782L...3C} and the theoretical expectation of a forming cluster \citep{2013ApJ...779..127C}. In the overdensity/field comparisons of galaxy populations in \S\,5 and \S\,7, we have considered these three $z=2.1$ LAEs located in large-scale overdensity.

\def\arraystretch{1.5}
\begin{centering}
\begin{deluxetable*}{cccccccccccc}
\tablecaption{HPS-COSMOS $\rm{Ly\alpha}$ Emitter Catalog in the $z=2.10$ structure} \tablewidth{0pt}
\tablehead{	\colhead{HPS}	&
		\colhead{$z$\tablenotemark{a}}	&
		\colhead{$\alpha$}	&
		\colhead{$\delta$}	&
		\colhead{Flux}	&
		\colhead{L}	&
		\colhead{Spectral}	&
		\colhead{Spatial}	&
		\colhead{Counter-} &
		\colhead{Counter-} &
		\colhead{EW$_{\rm{rest}}$$\rm$\tablenotemark{e}}	&
		\colhead{Flux$\rm{_{X-ray}}$}	\\
		\colhead{Index}	&
		\colhead{($\rm{Ly\alpha}$)}	&
		\colhead{(J2000)}	&
		\colhead{(J2000)}	&
		\colhead{($\rm{Ly\alpha}$)}	&
		\colhead{($\rm{Ly\alpha}$)}	&
		\colhead{FWHM$\rm$\tablenotemark{b}}	&
		\colhead{FWHM$\rm$\tablenotemark{c}}	&
		\colhead{part m$\rm{_R}$}	&
		\colhead{part Prob.$\rm$\tablenotemark{d}}	&
		\colhead{($\rm{Ly\alpha}$)}	&
		\colhead{(0.5--10 keV)}	\\
		\colhead{}	&
		\colhead{}	&
		\colhead{[deg]}	&
		\colhead{[deg]}	&
		\colhead{[$10^{-17}$ cgs]}	&
		\colhead{[$10^{42}$ cgs]}	&
		\colhead{[km s$^{-1}$]}	&
		\colhead{[arcsec]}	&
		\colhead{[mag]}	&
		\colhead{}	&
		\colhead{[\AA]}	&
		\colhead{[$10^{-17}$ cgs]}}
\startdata
\noalign{\vskip -1mm} 
244 & 2.0996 & 150.09858 & 2.22000 & 10.4$^{+5.2}_{-4.2}$ & 3.5$^{+1.8}_{-1.4}$ & 114 & 3.5$^{+1.5}_{-1.3}$ & 26.02 & 0.25 & 114.1$^{+89.3}_{-50.6}$ &  \\
261 & 2.0960 & 150.11904 & 2.29678 & 143.7$^{+23.2}_{-10.1}$ & 48.4$^{+7.8}_{-3.4}$ & 886 & 8.3$^{+0.9}_{-0.6}$ & 23.76 & 0.87 & 536.7$^{+157.8}_{-92.4}$ & 2040$\pm125$ \\
313 & 2.0975 & 150.16992 & 2.30656 & 25.1$^{+12.4}_{-10.1}$ & 8.5$^{+4.2}_{-3.4}$ & 249 & 5.0$^{+1.8}_{-1.3}$ & 22.75 & 0.98 & 23.9$^{+12.3}_{-9.7}$ & \\
\noalign{\vskip -3mm}
\enddata
\tablenotetext{a}{With an uncertainty of $4\times 10^{-4}$ based on a 0.5 \AA\ line center uncertainty.}
\tablenotetext{b}{After deconvolution with a 5 \AA\ FWHM instrumental resolution ($\sigma_{inst}\sim130$ km s$^{-1}$).}
\tablenotetext{c}{Including a tophat component of the fiber size of $4''.235$ and the effects of dither pattern and discrete sampling.}
\tablenotetext{d}{Probability of counterpart association ($R$-band).}
\tablenotetext{e}{Based on an interpolation between the two nearest filters for continuum.}

\end{deluxetable*}
\end{centering}

\end{document}